%% file: main-arxiv.tex
\include{Commands}
\documentclass[11pt]{article}
\usepackage[utf8]{inputenc}
\usepackage[english]{babel}
\usepackage{psfrag}

\usepackage{amsmath}
\usepackage{amsfonts}
\usepackage{amssymb}
\usepackage{bm}
\usepackage{xcolor}
\usepackage[version=4]{mhchem} 
\usepackage{mathtools}
\usepackage{amsbsy}
\usepackage{graphicx}
\usepackage{caption}
\usepackage{subcaption}
\usepackage[figurename=Figure]{caption}
\usepackage[tablename=Table]{caption}
\usepackage{cleveref}
\usepackage[square,sort,comma,numbers]{natbib}
\usepackage{geometry}
\usepackage{times}
\usepackage[color=green!20]{todonotes}

\usepackage{parskip}
\usepackage{color}
\usepackage{wrapfig}

\usepackage{siunitx}
\usepackage{adjustbox}
\usepackage{tabularx}
\renewcommand{\thetable}{\arabic{table}}
\usepackage{footnote}
\usepackage{threeparttable}
\usepackage{ulem}

\input{mathSymbols.tex}

\newcommand{\beginsupplement}{%
        \setcounter{table}{0}
        \renewcommand{\thetable}{S\arabic{table}}%
        \setcounter{figure}{0}
        \renewcommand{\thefigure}{S\arabic{figure}}%
     }
\title{\bfseries {A mechanical model reveals that non-axisymmetric buckling lowers the energy barrier associated with membrane neck constriction}}
\author{R. Vasan$^1$, S. Rudraraju$^2$, M. Akamatsu$^4$, K. Garikipati$^5$, and P. Rangamani$^{1*}$}

\begin{document}

\maketitle
$^1$Department of Mechanical and Aerospace Engineering, University of California San Diego, La Jolla CA 92093; $^2$ Department of Mechanical Engineering, University of Wisconsin-Madison, Madison, WI 53706, USA; $^3$ Biophysics Graduate Group, University of California, Berkeley, CA 94720, USA;  $^4$ Department of Molecular and Cell Biology, University of California, Berkeley, CA 94720, USA; and $^5$ Departments of Mechanical Engineering and Mathematics, Michigan Institute for Computational Discovery \& Engineering, University of Michigan, Ann Arbor, MI 48109, USA.
\\$^{*}$Corresponding Author\\
 $|$Email: padmini.rangamani@eng.ucsd.edu $|$

\maketitle

\begin{abstract}
Membrane neck formation is essential for scission, which, as recent experiments on tubules have demonstrated, can be location dependent. The diversity of biological machinery that can constrict a neck such as dynamin, actin, ESCRTs and BAR proteins, and the range of forces and deflection over which they operate, suggest that the constriction process is functionally mechanical and robust to changes in biological environment. In this study, we used a mechanical model of the lipid bilayer to systematically investigate the influence of location, symmetry constraints, and helical forces on membrane neck constriction. Simulations from our model demonstrated that the energy barriers associated with constriction of a membrane neck are location-dependent. Importantly, if symmetry restrictions are relaxed, then the energy barrier for constriction is dramatically lowered and the membrane buckles at lower values of forcing parameters. Our simulations also show that constriction due to helical proteins further reduces the energy barrier for neck formation compared to cylindrical proteins. These studies establish that despite different molecular mechanisms of neck formation in cells, the mechanics of constriction naturally leads to a loss of symmetry that can lower the energy barrier to constriction.
\end{abstract}

\fbox{\begin{minipage}{\textwidth}
\textbf{Significance statement} \\
Membrane tubule constriction is a critical step of cellular membrane trafficking processes and is thought to be mechanically regulated. Mechanical modeling techniques employing the Helfrich Hamiltonian and axisymmetric continuum frameworks have previously described energy barriers to constriction as a function of location along a membrane tubule. Recent advances in numerical modeling using spline basis functions (Isogeometric Analysis) enable us to conduct our analyses of membrane mechanics in a generalized 3D framework. Here, we implement a novel 3D Isogeometric Analysis framework and juxtapose it against an axisymmetric model to study the influence of location, symmetry constraints and helical collars on the constriction pathway. We show that an unsymmetric, ``crushed soda can" neck consistently displays a lower energy barrier than a symmetric neck.
\end{minipage}}

\newpage

\section*{Introduction}









Many cellular transport processes involving the plasma membrane including  different forms of endocytosis \cite{goldstein1979coated, rejman2004size, weinberg2012clathrin}, exocytosis \cite{sutton1998crystal, giraudo2006clamping}, and vesicle budding from intracellular organelles \cite{rothman1996protein, mathivanan2010exosomes} require mechanical deformation of the cellular membranes. The generation of membrane curvature is essential to trafficking, and the morphology of membranes has often been characterized as distinct shapes including U-and $\Omega$-shaped bud profiles \cite{avinoam2015endocytic, hassinger2017design, alimohamadi2018role} and tubulovesicular structures \cite{dippold2009golph3, kong2018cryo}. The molecular mechanisms of these processes can be attributed to biochemical components of the protein machinery involved \cite{mayor2007pathways, kaksonen2005modular}. For example, in the case of clathrin-mediated endocytosis (CME), more than 50 proteins are involved in regulating the different steps of membrane invagination such as nucleation, cargo selection, coat assembly, neck formation and scission \cite{kaksonen2018mechanisms, weinberg2012clathrin, engqvist2001actin} and contribute to the robustness and progression of endocytosis.

The formation of a membrane neck and scission are the last steps during many trafficking processes preceding vesicle formation. This neck formation is mediated by multiple biochemical mechanisms including mechanoenzymes belonging to the dynamin family \cite{macia2006dynasore}, helix insertion due to BAR domain proteins \cite{boucrot2012membrane} and ESCRT proteins \cite{hurley2010membrane, ramirez2019bidirectional}. A common organizational feature of these different proteins is that they form helical assemblies at the membrane neck through oligomerization \cite{ford2011crystal, guizetti2011cortical}.


Studies using reconstituted systems of lipid tubules decorated with protein assemblies have identified certain geometric and mechanical features of scission. Notably, studies of dynamin-mediated scission \cite{morlot2012membrane, roux2006gtp, danino2004rapid}, the most investigated scission mechanism, have shown that the location of neck formation along a membrane tube, membrane tension, and bending rigidity play important roles in membrane tube constriction and scission \cite{lenz2009mechanical}. Collectively, these studies support an emerging view that fundamental physical laws and geometric bounds confer a universality on membrane constriction phenomena and scission. 


Crucially, neck formation occurs at a length scale of $<$ 10 nm, which is challenging to image even with high resolution electron tomography (ET) as radiation damage and low signal to noise ratio (SNR) can limit contrast \cite{lidke2012advances}. Alternatively, equipped with extensive information from experiments such as those described above, mathematical and computational models can provide insight to the mechanics and energetics of membrane neck formation. Almost all of these models are rooted in the Helfrich elastic energy framework \cite{helfrich1973elastic}. The physical principles underlying the Helfrich model are simple enough -- the elastic energy of membrane deformation depends primarily on the curvatures of the membrane. Computational implementation of the governing equations resulting from this model, however, remain extremely challenging (see \cite{Guckenberger2017} for a detailed review). Therefore, many studies have assumed an axisymmetric configuration of the membrane for ease of computation \cite{chabanon2018gaussian, kozlov1999dynamin, kozlov2001fission, kozlov2010protein,kozlovsky2003membrane, hassinger2017design, alimohamadi2018role}.

\begin{figure}[b]
\centering
\includegraphics[width=0.85\linewidth]{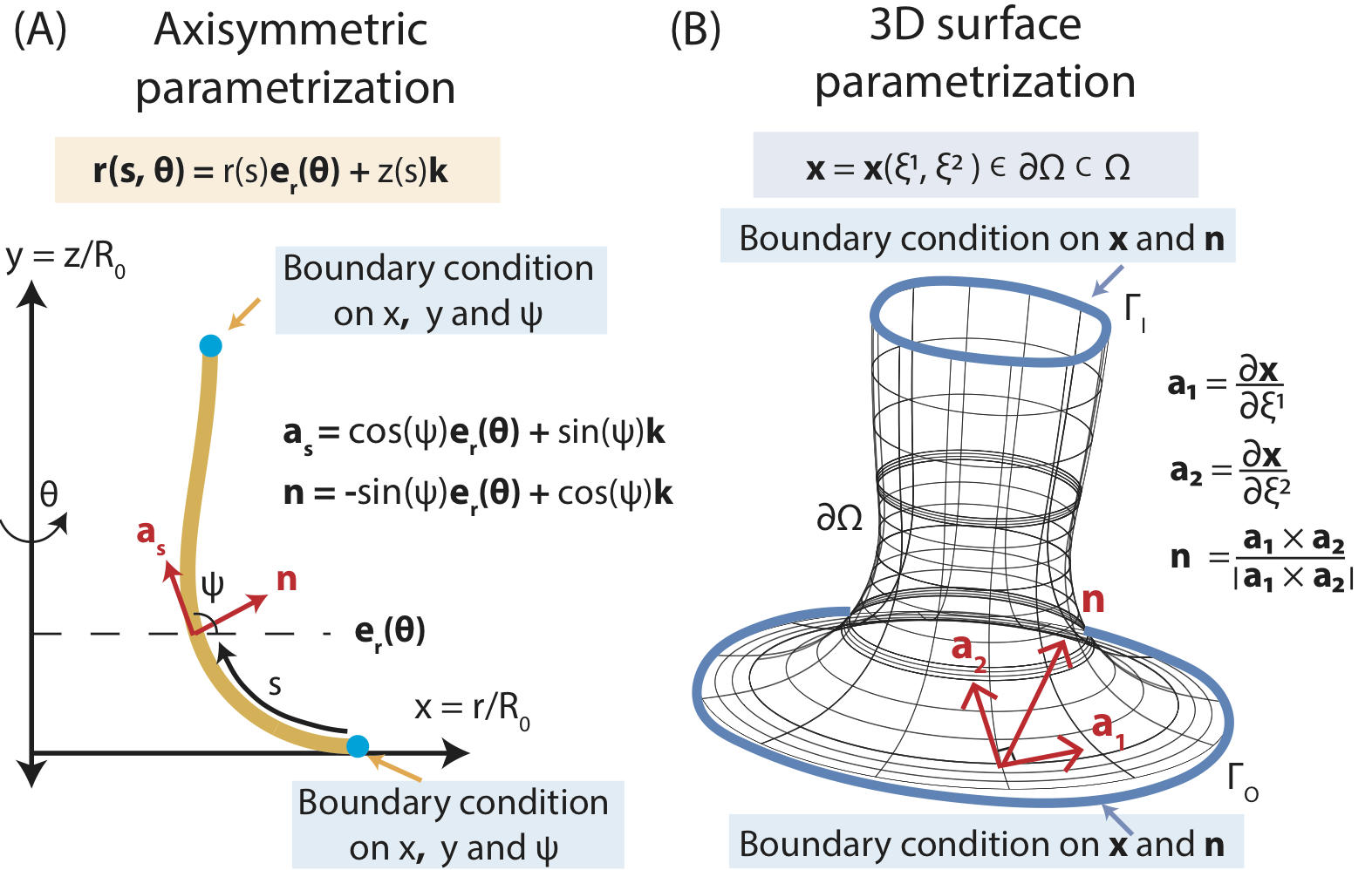}
\caption{Schematics showing surface parametrization of the membrane geometry in the axisymmetric and 3D formulations. (A) The axisymmetric coordinate system is parametrized in terms of the unit tangent vector ($\ba_{\bs}$), unit surface normal vector ($\bn$) and arc length ($s$), where $\br(s,\theta)$ is the position vector, $s$ is the arc length along the axisymmetric curve, $\theta$ is the out-of-plane rotation angle, r is the radius, z is the height, $\be_{\br}$ is the unit radial vector and $\bk$ is the unit axial vector. ($\be_{\br}$, $\be_{\theta}$, $\bk$) forms the coordinate basis (see SOM for more details). (B) Parametrization of a surface ($\partial \Omega$) embedded in a 3D volume ($\Omega$). Here, $\textbf{x}$ is the position vector of a point on the surface parametrized in terms of the surface coordinates ($\xi^1,\xi^2$) which are associated with a flat 2D domain that is then mapped to $\partial \Omega$ by $\textbf{x}=\textbf{x}(\xi^1,\xi^2)$. $\textbf{a}_\textbf{1}$ and $\textbf{a}_\textbf{2}$ are the local tangent vectors to the surface at $\textbf{x}$, and $\textbf{n}$ is the corresponding surface normal. ($\textbf{a}_\textbf{1}$, $\textbf{a}_\textbf{2}$, $\textbf{n}$) forms the local coordinate basis. The axisymmetric coordinate system in (A) is a specialization of the general curvilinear coordinate system depicted in (B).}
\label{fig:coordinates}
\end{figure}

In the most relevant of these studies to the present work, we and others have shown that a snap-through instability governs the first energy barrier associated with the formation of a membrane neck during CME \cite{hassinger2017design,walani2015endocytic, irajizad2019geometric}. An important limitation of the assumption of axisymmetry is that membrane deformation pathways associated with neck constriction that may have lower symmetries are not accessible (\Cref{fig:coordinates}B) and helicoidal protein assemblies \cite{martina2018role, hinshaw1995dynamin, kong2018cryo} cannot be explicitly modeled. 




In this study, we systematically investigate the energy barriers to constriction at different locations of a membrane geometry with and without symmetry restrictions (\Cref{fig:coordinates}). Importantly, we tackle the challenging problem of modeling non-axisymmetric membrane deformations with a benchmark comparison to axisymmetric modeling. We use a minimal, but fundamental, model of collar pressure-mediated tube constriction to obtain insights from a mechanical and energetic perspective. Using this model, we seek to answer the following foundational questions for the broader field of membrane deformation processes: \textit{First}, how does the local pre-existing curvature along a tube influence the energy barrier associated with neck constriction? \textit{Second,} how does relaxation of \textit{a priori} imposed symmetry restrictions impact the energy barriers associated with constriction of the neck?  \textit{And finally}, how do cylindrical versus helical protein assemblies modulate this energy barrier? 

\begin{figure}[t]
\centering
\includegraphics[width=0.85\linewidth]{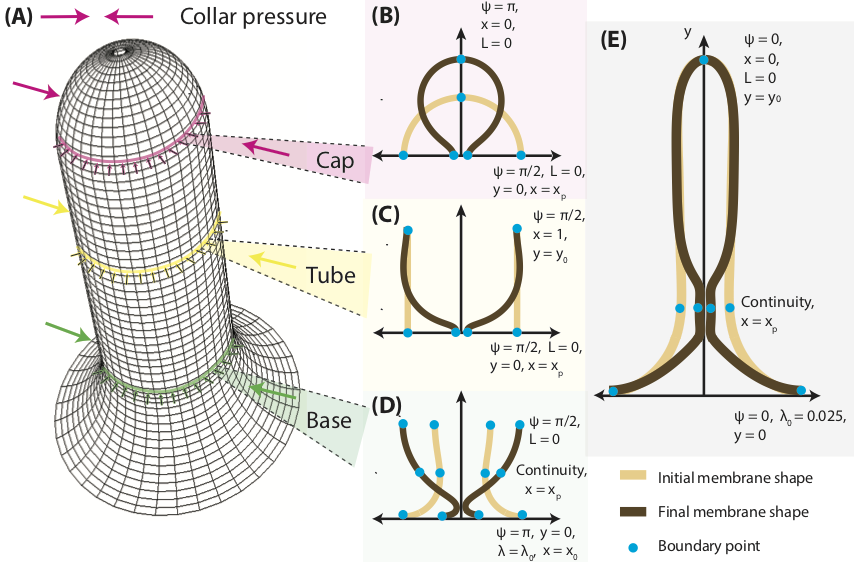}
\caption{Schematic depicting the modeling framework and simulation set up. Localized forces acting on the membrane were simulated as a collar pressure (A). Three different pinching locations are considered along the membrane tubule. (B) Case 1: Collar pressure applied at a circumference near the cap of the tube, where the mean and Gaussian curvatures are positive. (C) Case 2: Collar pressure applied at the center of the tube, where the mean curvature is positive and Gaussian curvature is zero. (D) Case 3: Collar pressure applied at a circumference at the base of the tube, where the mean curvature is positive along the cylindrical region and negative along the boundary, and Gaussian curvature is negative. (E) Case 4: Collar pressure applied along a tubule of fixed length pulled from an initially flat membrane. Shown are the initial membrane shape (light brown), final membrane shape (dark brown) and boundary/interface points (blue dots). x = x$_{p}$ is an interface condition enforced in axisymmetry to solve for the collar pressure as an unknown parameter. }
\label{fig:schematic}
\end{figure}

To answer these questions, we have developed a computational framework for solving membrane mechanics problems on complex geometries using numerical techniques that exploit Galerkin methods, specifically Isogeometric Analysis \cite{Cottrell2009}. This framework draws upon recent far-reaching advances on the use of spline basis functions in computational mechanics and brings them to the world of biological membranes, while building upon recent literature on finite element modeling of liquid shells \cite{sauer2017}. As a result, we can now investigate membrane deformation using simulations of neck constrictions under conditions that are notably less restrictive than those adopted previously in the literature (i.e. no enforced axis of symmetry). Importantly, this allows us to probe realistic helical constriction pathways within a continuum framework, a different approach than recent efforts using coarse-grained modeling \cite{martina2018role}. Using this framework, we applied constriction pressures at three different locations along the membrane tube (see \Cref{fig:schematic}) -- the `cap'(positive mean and Gaussian curvature), `cylindrical tube' (positive mean and zero Gaussian curvature) , and `base' (Positive- negative mean and negative Gaussian curvature). Our simulations show that the energy barriers associated with membrane neck constriction are indeed curvature-dependent, and therefore location-dependent, regardless of symmetry restrictions. Most importantly, we show that access to less symmetric shapes of membrane deformation lowers the energy barrier for scission considerably. These results suggest that loss of symmetry of the membrane neck may be an important energetic feature of successful neck formation.

\section*{Model development and simulations}
\subsection*{Helfrich energy}
The lipid bilayer is modeled as a thin elastic shell using the Helfrich energy \cite{helfrich1973elastic} based on the assumption that the thickness of the bilayer is negligible compared to its radius of curvature \cite{kishimoto2011determinants,morlot2012membrane}. The Helfrich energy density is defined as

\begin{align}
W = \kappa H^2 + \kappa_GK,
\label{general}
\end{align}

where $\kappa$ is the bending rigidity, H is the mean curvature, K is the Gaussian curvature and $\kappa_G$ is the Gaussian rigidity. Furthermore, we assume that the membrane is incompressible (\textit{i.e} the membrane area is constant ) \cite{evans1979mechanics}--a constraint that is implemented using a Lagrange multiplier field. Thus, while the Helfrich energy is defined entirely in terms of the geometry of the surface, the Lagrange multiplier, often interpreted as membrane tension \cite{rangamani2014protein,steigmann1999fluid}, is an important parameter that determines the minimum energy configuration. We ignore any fluid \cite{rangamani2013interaction, arroyo2009relaxation} and friction \cite{simunovic2017friction, quemeneur2014shape, rahimi2012shape}  properties of the bilayer, guided by the dominance of unstable and stable equilibrium states over relaxation/rate processes. The augmented Helfrich Hamiltonian that is being minimized on the surface $\Omega$, including the Lagrange multiplier $\lambda$ is given as \cite{steigmann1999fluid, agrawal2009modeling, rangamani2013interaction}

\begin{align}
E = \int_{\Omega} (\kappa H^2 + \kappa_GK+\lambda)dA.
\label{general}
\end{align}

\renewcommand{\arraystretch}{1.5}
\begin{table}
\centering
\caption{Model parameter values}
\begin{tabular}{lrrr}
Parameter & Value & Reference \\
\hline
1. Boundary membrane tension ($\lambda_{0}$) & 10$^{-2}$ - 10$^{-1}$ pN$\cdot$nm$^{-1}$ & \cite{mulholland1994ultrastructure, dai1998membrane} \\
2. Bending rigidity of bare membrane ($\kappa$) & 320 pN$\cdot$nm & \cite{dimova2014recent} \\
3. Length scale for non-dimensionalization ($R_0$) & 20 nm& \cite{hassinger2017design} \\
\hline
\end{tabular}
\label{table}
\end{table}
\subsection*{Simulations in axisymmetric coordinates}
In axisymmetry, the membrane is modeled using coordinates defined in \Cref{fig:coordinates}A. As the membrane tubule (Fig. \ref{fig:schematic}E) has three distinct shape features (Fig. \ref{fig:schematic}A - Cap, Tube and Base), local membrane geometries were modeled as a hemispherical cap (Fig. \ref{fig:schematic}B, Case 1), cylindrical tube (Fig. \ref{fig:schematic}C, Case 2) and a curved base with negative Gaussian curvature (Fig. \ref{fig:schematic}D, Case 3). Cases 1 and 2 are constant mean curvature shapes and are solved as two-point boundary value problems. Case 3 is a negative Gaussian curvature shape with an inflection point in mean curvature with respect to the arc length and is solved as a three point boundary value problem. Case 4 includes local geometric variations in both mean and Gaussian curvature and is solved as a three point boundary value problem. The third point in these cases is an additional interface point enforced at the location of constriction, such that it satisfies continuity requirements \cite{shampine2000solving}. The resulting system of equations is solved using the partial differential equation solution routines in Matlab, specifically bvp4c \cite{shampine2000solving, venkataraman2009applied}. Importantly, these equations are solved using both force control (compute membrane shape for a certain applied force) and displacement control (compute applied force for a certain membrane shape). These two approaches can lead to the same equilibrium membrane shape. In the presence of membrane bending instabilities, displacement control can access regimes of the response curve that force control cannot reach. However, this requires a precise prescription of the kinematic path. In order that a system be free to find the lowest energy pathways through a region of instability in its energy landscape, it is important that, while the force and displacement vary in a coupled manner, neither quantity be fully prescribed \cite{agrawal2009boundary,hassinger2017design}. Parameters for the bending rigidity, membrane tension and non-dimensionalization length R$_0$ are specified in Table 1. Details of the numerical methods are provided in the Supplementary Online Material (SOM).

\subsection*{3D numerical model development and validation}


The membrane deformation problems considered in this paper can be modeled using classical thin shell theories of mechanics. However, given the geometric complexity and the associated boundary conditions, analytical solutions are inaccessible. Instead, we obtain three dimensional numerical solutions to the membrane deformation problems using the framework of Isogeometric Analysis (IGA) \cite{Cottrell2009}. An IGA method-based membrane mechanics framework has been developed for this work, and is build on top of the PetIGA \cite{2016petiga} open source library. In an IGA approach, the membrane geometry is discretized using a spline mesh and the governing equations (Fig. 1B, see thin shell formulation in the Supplementary Information) are converted to a nonlinear system of equations. This nonlinear system of equations is then solved to obtain the deformed membrane shape, and the related force and energy metrics. Of importance to our central result is that this framework naturally admits both symmetric and asymmetric deformation modes driven by the underlying physics. This framework has three key assumptions. First, a fundamental conjecture of the Helfrich model is that the characteristic length scales of the problem are much larger than the thickness of the bilayer \cite{helfrich1973elastic}. This assumption allows us to neglect the effect of transverse shear deformations and consider the classical Kirchhoff–Love shell kinematics for thin shell geometries \cite{novozilov1959}. Second, numerical solutions to the membrane shape equations (\Cref{eq:s-normal,eq:s-Tangential}) in general coordinates are challenging because of continuity requirements in the numerical scheme. 
We have overcome this challenge by adopting both B-Spline basis functions, which allow high-order continuity, and the numerical framework of Isogeometric Analysis \cite{Cottrell2009}. Finally, an inherent limitation of the Helfrich energy formulation in three dimensional simulations is the lack of resistance to shear deformation modes. The zero energy modes corresponding to shear deformation are eliminated in this framework by adding shear stabilization terms of smaller magnitude relative to the traditional bending terms in the Helfrich energy \cite{sauer2017}, thus restoring stability to the numerical model. A companion manuscript (in preparation by the authors) describes the details of the mathematical methods and numerical formulation, and establishes the validity of the computational framework by modeling a range of problems in membrane mechanics. Here, we present a validation of the 3D computational framework by comparing the output from the simulation with a known analytical solution of the classical tube pulling problem (\Cref{fig:pullout_comparison}A, B). In addition to demonstrating good agreement with the analytical solution, the 3D model also resolves the symmetric pathways of deformation if they are indeed the energy minimizing modes (\Cref{fig:pullout_comparison}A). 
Having validated the 3D numerical scheme, we then proceeded to simulate the different cases shown in (\Cref{fig:schematic}B-E) and compared them against axisymmetric pathways. We use three key metrics to compare the two models -- (1) the radial pinching load, represented by the collar pressure that drives constriction, (2) structural stiffness of the membrane, defined as the slope of the load-displacement response, and (3) membrane bending energy. We track these metrics for different pinching radii, which are defined as the shortest distances between the membrane and the center of the necking region. For fully symmetric configurations and those with lower symmetry, this distance is the radius of the smallest circle that can be fit in the necking region. 

\section*{Results}
We systematically investigated the role of preexisting curvature (varying with location on the membrane) in the constriction process and the associated energy landscape using both the traditional axisymmetric calculations and the 3D computational framework. The constriction process is modeled using a collar pressure (pN/nm$^2$) applied onto a fixed membrane height (nm). In this study, we include the effect of the height of applied pressure by reporting a force per unit length, or an effective surface pressure (pN/nm), as the product of the applied collar pressure and the fixed height. Our main results can be summarized as follows -- \textit{first}, the energy landscape for constriction depends on the preexisting curvature of the membrane; \textit{second}, 3D modes of constriction with less than full symmetry encounter lower energy barriers when compared to pathways of higher symmetry; and \textit{finally}, helical constriction modes can have the lowest energy barriers of all in 3D. We elaborate on these findings in detail below.   




\subsection*{The energy barrier associated with constriction depends on preexisting membrane curvature}
We investigated the effect of local, preexisting curvature on the energy barrier associated with tubule constriction in axisymmetry. We pulled out a membrane tube by applying an external axial force (f$_{axial}$) on a small patch of the membrane to mimic a point load while maintaining a membrane tension of 0.2 pN/nm \cite{derenyi2002formation}. We then applied a radial collar pressure at different locations on the tube (\Cref{fig:fig_asymm_whole_tube}A) while maintaining the membrane height, a setup that can be generalized to \textit{in-vitro} membrane tubules pulled by optical tweezers. In the absence of a fixed height applied as a boundary condition, the membrane deforms freely in the axial direction at negligible collar pressures (\Cref{fig:S1_boundary_conditions}). Results from our simulations show that pinching the tube at the cap (positive mean and Gaussian curvature) and along the cylinder (positive mean and zero Gaussian curvature) results in similar force-shape relationship (\Cref{fig:fig_asymm_whole_tube}D) and the cross section of the pinched profile remains circular by construction due to the restriction of axisymmetry. (\Cref{fig:fig_asymm_whole_tube}B,C). Surprisingly, for the same range of collar pressure applied to the base (positive-negative mean and negative Gaussian curvature), we observed the existence of a snap-through instability as the membrane constriction progresses, as shown by the red line in \Cref{fig:fig_asymm_whole_tube}D. The dotted lines (\Cref{fig:fig_asymm_whole_tube}D, base) are calculated using displacement control, i.e., compute the applied force given the membrane shape. However, given our initial conditions and the mechanism of neck formation via increasing pressure, these shapes are not accessible during constriction.  
As in all snap-through instabilities, this pinching instability arises from a reduced energy barrier and associated reduction in neck radius, and has been reported in other membrane physical processes as well \cite{hassinger2017design, walani2015endocytic, irajizad2019geometric}. Despite the existence of the snap-through instability at the base, the pressure needed for further constriction becomes unbounded as the pinching radius approaches zero. This suggests that fully symmetric membrane shapes are not favorable for constriction below a certain critical radius. 


\subsection*{Relaxation of symmetry constraints lowers the energy barrier associated with membrane constriction}
\begin{figure}[!!h]
\centering
\includegraphics[width=\linewidth]{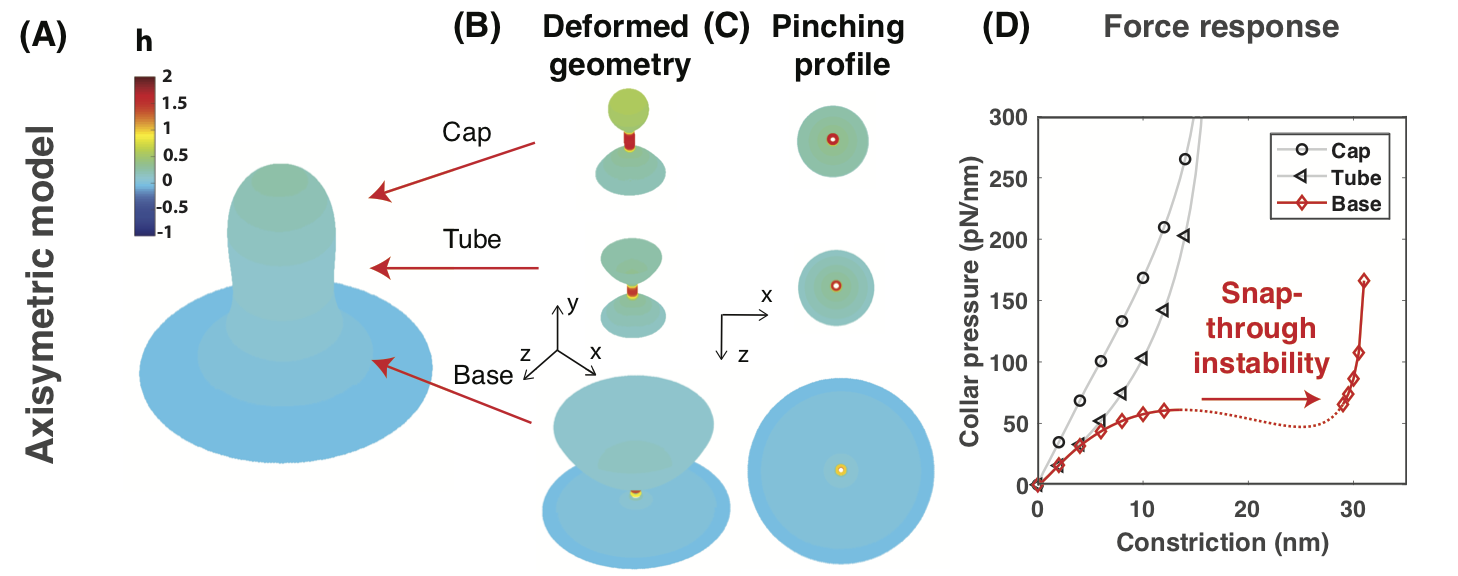}
\caption{Location dependence of membrane tube constriction in axisymmetry. Shown are the different locations of constriction, cap, tube and base (A), the corresponding membrane shapes (pinched configurations) obtained (B, C) and the evolution of the collar pressure as a function of constriction (force response, D). The bending rigidity $\kappa$ is 320 pN-nm, membrane tension $\lambda$ is 0.2 pN/nm, radius of the tube is 20 nm, the radially inward-directed collar pressure is applied over a strip of height 1 nm, and the height of the membrane tubule is 100 nm. The cap and the tube locations deform smoothly, while a snap-through instability is observed at the base  (D). The dotted solution path is never realized during the loading phase, leading to a transition to a wider tube morphology that is markedly different from the other cases. The colorbar in (A) shows the non-dimensional mean curvature.}
\label{fig:fig_asymm_whole_tube}
\end{figure}

\begin{figure}[!!h]
\centering
\includegraphics[width=\linewidth]{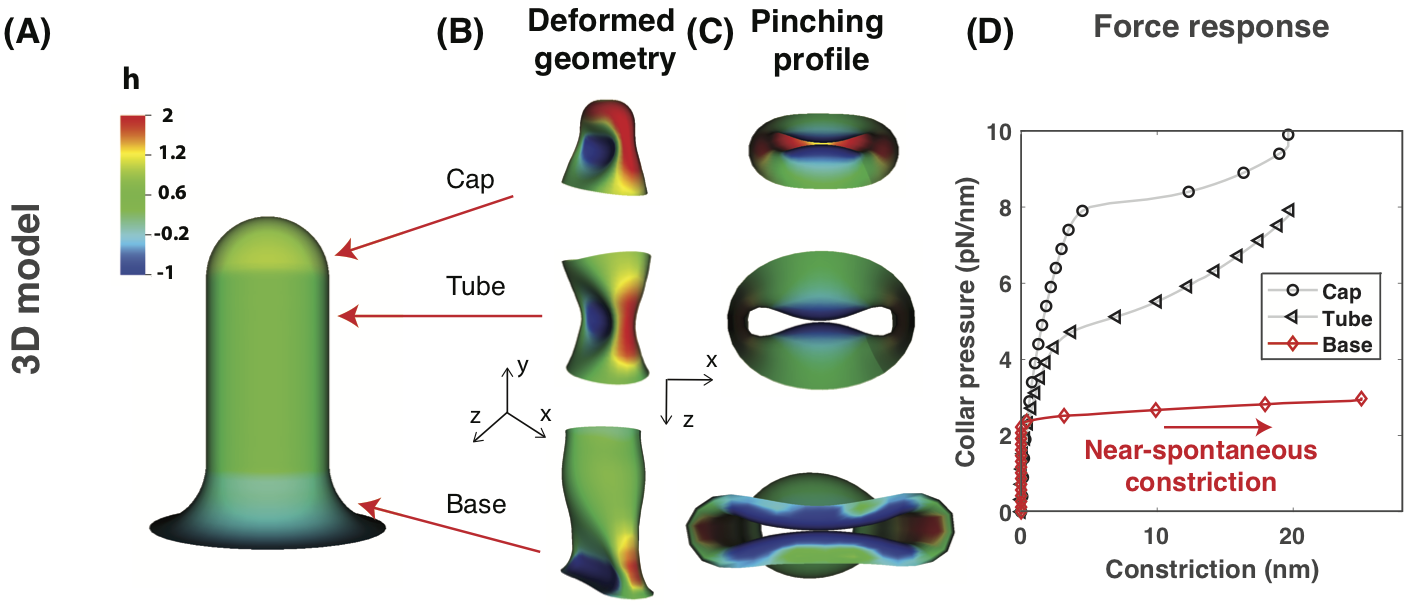}
\caption{Location dependence of membrane tube constriction in 3D. Shown are the different locations of constriction: cap, tube and base locations (A), the corresponding membrane shapes (pinched configurations) obtained (B, C) and the evolution of the collar pressure as a function of constriction (force response, D). Bending rigidity $\kappa$ is 320 pNnm, membrane tension $\lambda$ is 0.2 pN/nm, radius of the tube is 20 nm, height of the applied force is 1 nm, height of the membrane tubule is 100 nm. A near-spontaneous collapse is observed for the base, and a relatively stable constriction evolution for the cap and the tube locations (D). The colorbar in (A) indicates non-dimensional mean curvature. See Movies M1-M3 in the supplementary information for the evolution of the constriction process for the Cap, Tube and Base locations.}
\label{fig:fig_3D_whole_tube}
\end{figure}

We next asked if relaxation of symmetry constraints alters the energy landscape of location-dependent constriction. 
To answer this question, we used our 3D model. For these simulations, we initialized the computation as a pre-formed membrane tubule to limit computational complexity (see SOM). 
Strikingly, we observed that once the symmetry constraints are relaxed, membrane constriction at all three locations requires a lower collar pressure by more than an order of magnitude when compared to the axisymmetric deformation (compare \Cref{fig:fig_3D_whole_tube}D and \Cref{fig:fig_asymm_whole_tube}D). To verify this result, we enforced axisymmetry constraints in the 3D model and  repeated our calculations for the ``cap" (\Cref{fig:schematic}B) and ``tube" (\Cref{fig:schematic}C). We observed that the collar pressures increased by an order of magnitude (\Cref{fig:disp_control_cap_pinching,fig:disp_control_tube_pinching}) when symmetry is imposed, resulting in comparable pinching profiles between the axisymmetric and 3D models. However, without the imposed  axisymmetry, the collar pressures reduced significantly (\Cref{fig:force_control_tube_pinching}). These results allow us to conclude that absence of enforced axisymmetry alone is responsible for the significant decrease of collar pressure. We next analyzed the shapes of the membrane cross sections during 3D constrictions, which we found to be distinctly reminiscent of buckling phenomena that are observed in thin walled elastic structures \cite{Azzuni2018, Rahman2011}. The pinching profiles shown in \Cref{fig:fig_3D_whole_tube}C correspond closely to the classical result of the first buckling mode of a thin ring subjected to inward pressure on its walls. These deformation modes of buckling/pinching in thin-walled elastic rings and tubes have been known since the early twentieth century in the context of structural engineering applications \cite{von1936kritische, windenburg1934collapse, wah1967buckling, kozlovsky2014general, hazel2017buckling, azzuni2019behavior}, and are also observed experimentally \cite{ross2008inelastic}. 




Finally, we observed that the base, with the preexisting negative Gaussian curvature, needed lower collar pressure to undergo constriction. This result is consistent with the observation that membranes with a negative Gaussian curvature are more amenable to constriction. Furthermore, contrary to the case of axisymmetric deformation, the collar pressure associated with increasing constriction at the base does not continue to grow with constriction (compare \Cref{fig:fig_3D_whole_tube}D, red line and \Cref{fig:fig_asymm_whole_tube}D, red line). While the collar pressure increases sharply for initial constriction (\Cref{fig:fig_3D_whole_tube}D, inset), it only rises gradually for the tube and cap geometries as constriction continues to increase several-fold. The initial increase in compressibility represents the natural stiffness of the membrane, after which symmetry breaking occurs leading to near spontaneous constriction. This result suggests that without the arbitrary restriction to axisymmetric deformation, near-spontaneous collapse of the neck is possible after a critical collar pressure is reached.  The sudden drop in stiffness observed for the base geometry (\Cref{fig:fig_3D_whole_tube}D) is associated with symmetry-breaking. As shown, this symmetry-breaking and loss of stiffness is not immediate, but occurs after a small amount of constriction has occurred. Thus, we predict that the energy landscape of membrane neck constriction is location dependent, but more importantly, predict that lower symmetry shapes attained by 3D constriction can significantly lower the energy barrier at the base to promote easier constriction. 

\begin{figure}[!!h]
\centering
\includegraphics[width=0.8\linewidth]{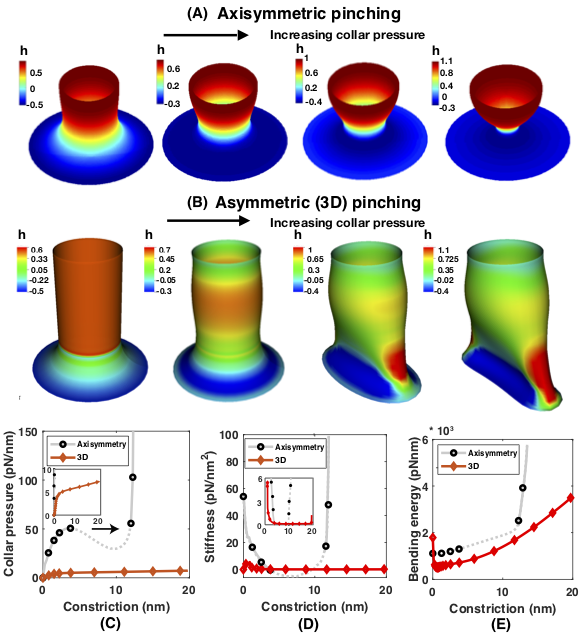}
\caption{Pinching response of the base geometry and the corresponding evolution of pressure, stiffness and bending energy obtained using the axisymmetric and 3D models. Boundary conditions are shown in Fig. 1, Case 3. Bending rigidity $\kappa$ is 320 pN$\cdot$nm, membrane tension is 0.2 pN/nm, the collar pressure is applied over a height of 1 nm, and  the length scale is set by the initial radius of 20 nm. Shown are the membrane shape evolution obtained from the  axisymmetric (A) and 3D models (B), and the corresponding variation of the collar pressure (C, with inset), stiffness (D, with inset), and bending energy (E). Colorbars in (A) and (B) indicate non-dimensional mean curvature. See Movie M3 in the supplementary information for the evolution of the constriction process for the base geometry.}
\label{fig:comparison_base_pinching}
\end{figure}

\begin{figure}[!!h]
\centering
\includegraphics[width=\linewidth]{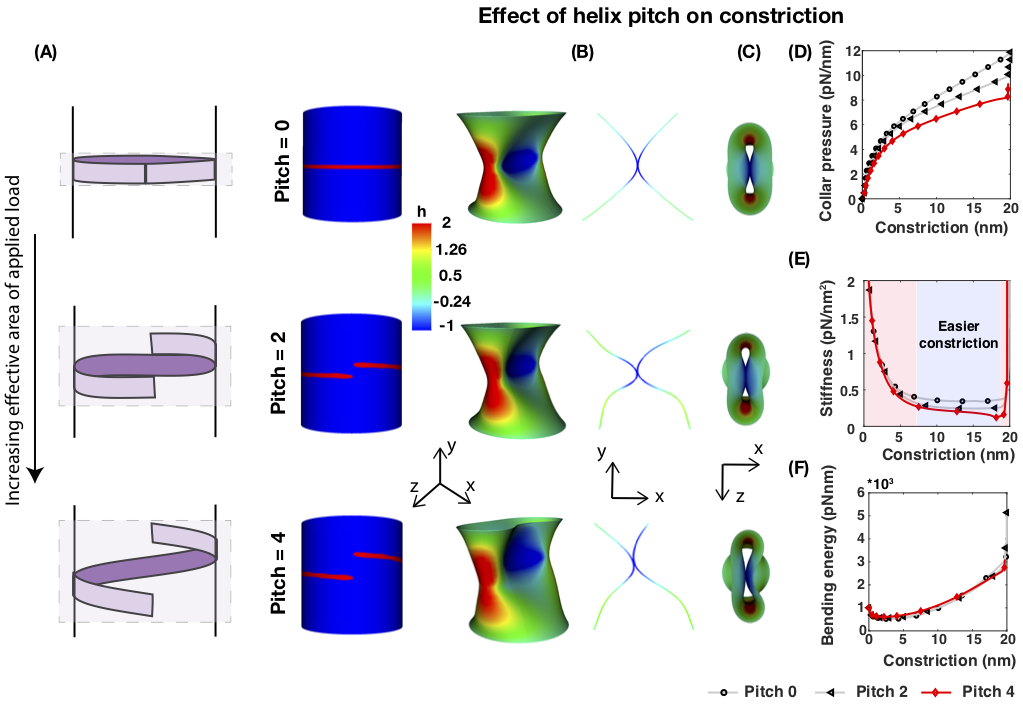}
\caption{A helical force collar further reduces the barrier to constriction. Considering force collars with a normalized pitch of zero, two and four, shown are the shape of the helical collar (A) with a schematic depicting increased span of load distribution area with increasing helical pitch, deformed shape and corresponding pinching shape (B), the pinching profile (C), and evolution of the collar pressure (D), stiffness (E), and membrane bending energy (F). Shaded regions of pink and blue in (E) represents a region of high stiffness ($> \sim$ 0.25 pN/nm$^2$) and low stiffness ($< \sim$ 0.25 pN/nm$^2$) respectively. The colorbar under (B) indicates non-dimensional mean curvature. See Movies M4-M6 in the supplementary information for the evolution of the constriction process due to a helical force collar at the tube location with a non-dimensional pitch of zero, two and four, respectively. }
\label{fig:fig_helix}
\end{figure} 

\begin{figure}[!!h]
\centering
\includegraphics[width=\linewidth]{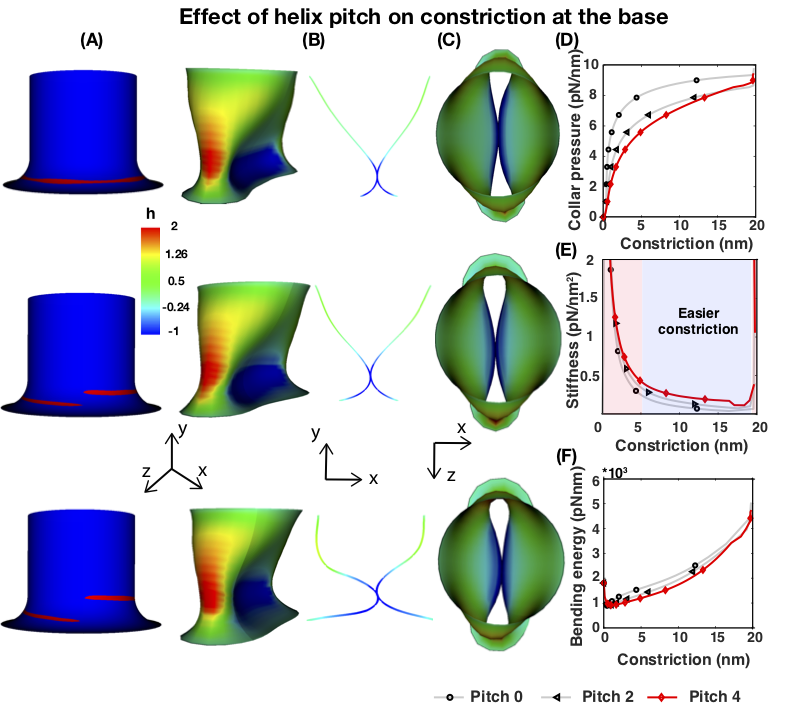}
\caption{A helical force collar increases stiffness to constriction at the base. Considering force collars with a normalized pitch of zero, two and four, shown are the  shape of the helical collar and deformed shape for pitch two and four (A-C), evolution of the collar pressure (D), stiffness (E), and membrane bending energy (F). Shaded regions of pink and blue in (E) represents a region of high stiffness ($> \sim$ 0.25 pN/nm$^2$) and low stiffness ($< \sim$ 0.25 pN/nm$^2$) respectively. The colorbar under (B) indicates non-dimensional mean curvature. See Movies M8-M10 in the supplementary information for the evolution of the constriction process due to a helical force collar at the base location with a non-dimensional pitch of zero, two and four, respectively.} 
\label{fig:fig_helix_base}
\end{figure} 

\begin{figure}[!!h]
\centering
\includegraphics[width=\linewidth]{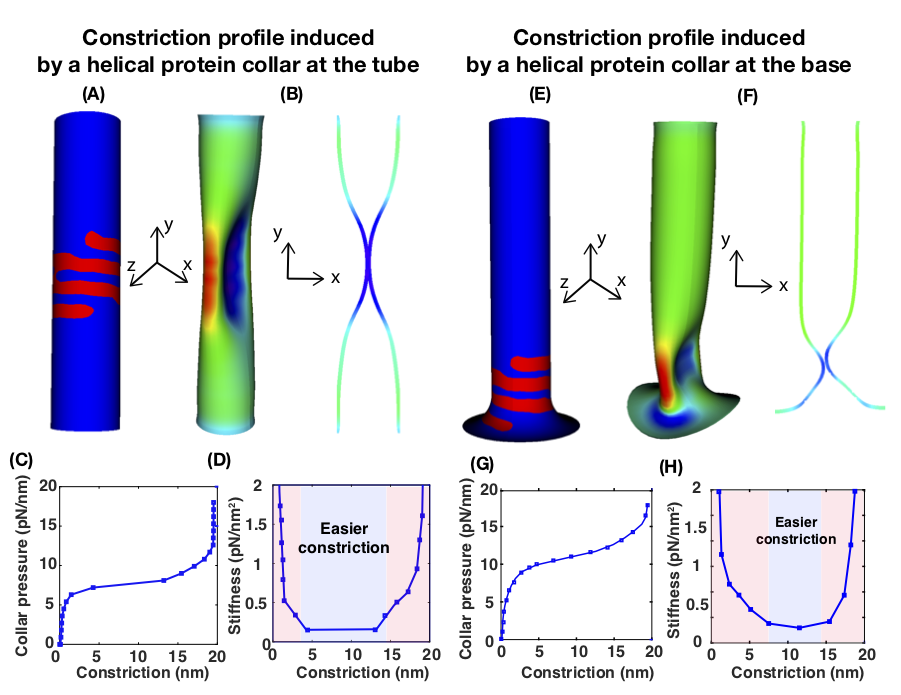}
\caption{Constriction by multiple helical rings are also location dependent. For both the tubule and base geometry, shown are the initial geometry and location of the force collar with three helical rings (A, E), the deformed shape and corresponding pinching profile (B, F) and the corresponding variation of collar pressure (C, G) and stiffness (D, H). See movies M7 and M11 for the corresponding evolution of the constriction process due to a force collar with three helical rings.} 
\label{fig:fig_multiple_helix}
\end{figure} 

\subsection*{Easier constriction at the base is accompanied by reduced membrane stiffness.}
We further investigated the energy landscape at the base of the tubule (\Cref{fig:schematic}D) to identify the mechanisms associated with easier constriction. 
For the axisymmetric pinching pathway, an increase in collar pressure results in progressive transformation of the tubule into a half-catenoid-shaped membrane. Indeed, this is the shape that  is commonly seen in schematics of membrane pinching (\Cref{fig:comparison_base_pinching}A). On the other hand, application of increasing collar pressure in 3D demonstrates that the membrane base is quick to break symmetry, and assumes the iconic shape of a soda can crushed by radial pinching (\Cref{fig:comparison_base_pinching}B).
We thus observe that the axisymmetric and 3D models invoke different constriction pathways; the axisymmetric model yields uniform pinching  (\Cref{fig:comparison_base_pinching}A), but the 3D model captures an asymmetric flattened tubule geometry (\Cref{fig:comparison_base_pinching}B).


As before, the axisymmetric mode shows a snap-through instability (\Cref{fig:comparison_base_pinching}C, \Cref{fig:fig3_axisymmetric_base}B). This instability can be attributed to a build up of negative tangential stress or tension work (\Cref{fig:fig3_axisymmetric_base}E). The sharp increase of both the bending energy (\Cref{fig:fig3_axisymmetric_base}D) and tension work (\Cref{fig:fig3_axisymmetric_base}E) at large constriction corresponds to a sharp increase in the pressure required for constriction (\Cref{fig:comparison_base_pinching}C, \Cref{fig:fig3_axisymmetric_base}B). However, in 3D, a narrow constriction radius is accessible at a much lower pressure when compared with the axisymmetric mode. This result can be understood by analyzing the relationship between stiffness and constriction of the membrane in different modes of deformation   (\Cref{fig:comparison_base_pinching}D). The stiffness of the membrane is significantly reduced in the lower symmetry mode attained in 3D when compared with the axisymmetric mode (compare black circle lines and red diamond lines in \Cref{fig:comparison_base_pinching}D). Similar comparisons for the tube (\Cref{fig:force_control_tube_pinching}) show a significantly reduced stiffness in lower symmetry pathways of 3D constriction when compared with axisymmetric pathways. Comparison of the bending energy in the axisymmetric and 3D modes of deformation shows that while the bending energies in both cases are similar in magnitude, the energy landscape is different (\Cref{fig:comparison_base_pinching}E). All intermediate energy states along the constriction of the neck radius are accessible in the 3D pathway of deformation while in the axisymmetric pathway the energy states associated with the snap-through regime are inaccessible (dashed grey region in \Cref{fig:comparison_base_pinching}E). From these analyses, we conclude that easier constriction at the base of the tube, revealed by full 3D computations, is accompanied by a reduced membrane stiffness and accessibility to all intermediate energy states along the constriction path. 

\subsection*{Helical force collar further reduces the barrier to membrane neck constriction}

The 3D model of membrane deformation allows us to probe the response of the membrane to non-symmetric force distributions such as those exerted by helical arrangements of proteins that cannot be modeled in the axisymmetric framework. Membrane scaffolding proteins such as dynamin \cite{kozlov2010protein, hinshaw1995dynamin} and ESCRT-III \cite{Hurley2010} self-assemble into helical collars that can constrict the neck. Recent Cryo-EM maps of human dynamin-1 (dyn-1) polymer report detailed structural and molecular information on its helical geometry \cite{kong2018cryo}. With GTP hydrolysis, the helical polymer actively constricts the membrane from a diameter of more than 20 nm to below 3.4 nm \cite{kong2018cryo}. Disassembly of dynamin is thought to promote scission via hemifission \cite{martina2018role, Bashkirov2008, mattila2015hemi}. Despite the well-established mechanisms of dynamin-mediated constriction, the response of the membrane to constriction and subsequent scission is not yet fully understood. This led us to investigate the role of a helical collar pressure, which is a mimic of force generated by helical protein assemblies, on membrane constriction. We explore the role of two geometric parameters of a helical collar -- the pitch, defined as the distance along the axis for a complete helical turn, and the number of rings. 

 We first consider a single helical ring exerting a collar pressure on a cylindrical tube (\Cref{fig:fig_helix}A) with different values of the pitch, non-dimensionalized by the height of the collar. Increasing helical pitch corresponds to an increased span (with same collar area) over which the load is distributed on the geometry. Qualitatively, we observe that the cross section of the neck is non-axisymmetric for different values of the pitch, confirming that the lower symmetry modes of deformation are still preferred for neck constriction with helical rings of pressure (\Cref{fig:fig_helix}B, C). Quantitatively, the collar pressure associated with constriction decreases as the pitch increases (\Cref{fig:fig_helix}D). Correspondingly, the stiffness also decreases for increased pitch (\Cref{fig:fig_helix}E), indicating that the bending energy (\Cref{fig:fig_helix}F) becomes slightly less steep. The pink and blue shaded regions in \Cref{fig:fig_helix}E indicate regions of high and low stiffness (easier constriction) respectively. From these observations, we conclude that helical collars have the ability to further reduce the energy barrier to membrane constriction by a decrease in collar pressure and stiffness associated with this process, and that the pitch of the helix is an important determinant of this barrier. 
 
To our knowledge, this is the first numerical characterization of the effect of helical squeezing forces on membrane constriction in a continuum framework. However, from a soft matter perspective, it is well-known that helical structures are known to exert squeezing forces more effectively on their support; an excellent example of this behavior can be found in the twining of plant vines \cite{goriely2006mechanics, isnard2009tensioning, kempaiah2014nature} and other naturally occurring soft materials. Like cylindrical forces, helices also exert tangential and radial forces. Additionally, because of the pitch of the helix, they also exert axial forces.
 


Given our observation that the base of the tube is energetically favorable to constriction (\Cref{fig:comparison_base_pinching}), we next added a helical collar pressure to the base to investigate the effect of coupling the three key design elements -- negative Gaussian curvature, access to non-axisymmetric modes, and helical collar pressure (\Cref{fig:fig_helix_base}A) -- on membrane neck constriction. We found that this combination also results in reduced collar pressure (\Cref{fig:fig_helix_base}D) and energy barrier (\Cref{fig:fig_helix_base}F) with increasing constriction when compared to a ring of collar pressure (\Cref{fig:fig_helix_base}D,F for zero pitch). However, a ring of lower pitch was advantageous in terms of a lower stiffness; the stiffness for Pitch 4 is greater than that associated with Pitch 0 (\Cref{fig:fig_helix_base}E).  This is possibly the result of recruiting the stiffer cylindrical portion of the membrane tube with greater pitch.  

Comparing the effect of helical pinching on the tube (\Cref{fig:fig_helix}) versus the base (\Cref{fig:fig_helix_base}), we arrive at the following conclusions. For a single helical ring, the relationship between helix geometry and the membrane geometry is non-trivial. It appears while both the tube and base geometry show a clear relationship between helical pitch and collar pressure (\Cref{fig:fig_helix}D, \Cref{fig:fig_helix_base}D), the same is not true for the stiffness (\Cref{fig:fig_helix}E, \Cref{fig:fig_helix_base}E). This suggests a complex interaction between the geometries of the membrane and the helical ring, possibly due to a mechanical feedback between membrane curvature and the constricting action of the helicases.



\subsection*{Effects of multiple helical collars are also location-dependent}

Since a helical ring of collar pressure offers a lower energy barrier to constriction, we next asked if  an increase in the number of helical rings can further promote scission. This question is motivated by observations that an increase in the number of dynamin rings is a predicted response to delayed scission and higher membrane tension \cite{grassart2014actin, roux2006gtp, shnyrova2013geometric, chappie2011pseudoatomic}. 
 To answer this question, we simulated 3 rings of helical collar pressure reminiscent of dynamin rings assembled on a membrane tubule \cite{shnyrova2013geometric, hinshaw1995dynamin, kong2018cryo}. More than 3 rings of a dynamin collar are unlikely to exist \textit{in vivo} due to disassembly of the dynamin oligomer \cite{pawlowski2010dynamin}. Collar pressure due to multiple helical rings not only leads to membrane constriction as expected (\Cref{fig:fig_multiple_helix}A) but also appears to stabilize the membrane tube against a sideways wobble that is observed with one ring (compare \Cref{fig:fig_helix}B with \Cref{fig:fig_multiple_helix}B). We also observed that while the values of the collar pressure are of the same order of magnitude for 3 rings as with 1 ring (compare \Cref{fig:fig_helix}D and \Cref{fig:fig_multiple_helix}C), the stiffness profile is different (\Cref{fig:fig_multiple_helix}D). During the early part of the constriction, the stiffness values go from high to low, quickly leading to a region of easier constriction (\Cref{fig:fig_multiple_helix}D, blue shaded region). However as constriction increases, the stiffness increases again over a finite range of constriction and at a lower value of constriction (\Cref{fig:fig_multiple_helix}D, pink shaded region), unlike the very narrow range of stiffening in \Cref{fig:fig_helix}E. This may be due to the increase in the surface area that is constricted by three rings as compared to the surface area constricted by one ring, increasing the structural resistance to constriction. Thus, multiple rings assist neck formation on a tubule during an initial constriction region (\Cref{fig:fig_multiple_helix}D, 4 - 15 nm of constriction), after which disassembly and possibly additional proteins are required.

 Interestingly, the presence of 3 rings at the base resulted in an increase in both the collar pressure (\Cref{fig:fig_multiple_helix}G) and stiffness (\Cref{fig:fig_multiple_helix}H), such that the region of easier constriction (\Cref{fig:fig_multiple_helix}H, blue shaded region) is much smaller than for a single helical ring (\Cref{fig:fig_helix_base}E, blue shaded region). The membrane then transitions into a region of high stiffness at a smaller value of constriction (\Cref{fig:fig_multiple_helix}H, pink shaded region) due to the larger surface area of the three rings that recruits more of the cylindrical tube to resist constriction.
 
  Furthermore, multiple helical rings achieve easier constriction at shorter constriction distances for a cylindrical geometry (\Cref{fig:fig_multiple_helix}D, blue region)  and at larger constriction distances for the base geometry (\Cref{fig:fig_multiple_helix}H, blue region). However, they resist further constriction at narrow radii independently of the pre-existing curvature (\Cref{fig:fig_multiple_helix}D and \Cref{fig:fig_multiple_helix}H, pink region) . Given these observations, it is possible that helical polymers might preferentially undergo conformational rearrangements such as a change in pitch or number of rings based on feedback with the underlying membrane curvature so as to achieve a lower energy barrier to constriction. Such structural rearrangements in dynamin have also been reported in experiments \cite{kong2018cryo}.

\section*{Discussion}


Membrane constriction and subsequent scission are universal to membrane remodeling processes \textit{in vitro} and \textit{in vivo}. While the molecular machineries may differ across systems, these deformation processes likely share the same common physical principles. In this study, using computational modeling, we show that there are three key design elements that play important roles in promoting membrane constriction -- (1) location \textit{i.e.} preexisting curvature of the membrane being constricted, (2) access to lower-symmetry modes of deformation, and (3) access to helical loading. 

From a mechanical standpoint, membrane constriction can be interpreted as a deformation mechanism driven by a radial collar pressure applied by the scission proteins in the vicinity of the necking region. For axisymmetric constriction, the pinching pressure needed to cause membrane constriction increases with the narrowing of the neck radius. This monotonic growth of the radial pressure results in a high energy barrier for pinching. Interestingly, many elastic structures have inherent modes of instability that result in enhanced deformation or even collapse in response to loading and are associated with lower energy barriers. Such modes are ubiquitous in thin elastic shells and manifest as folding, wrinkling, creasing, and buckling deformations (e.g. wrinkling of thin membranes and graphene sheets \cite{Deng2016}, surface tension induced buckling of liquid-lined elastic tubes \cite{hazel2005surface}, snap-through of elastic columns \cite{brojan2007buckling}, barrelling modes of thin cylinders \cite{Azzuni2018, Rahman2011}, etc.). Notably, they have lower symmetry than the fully axisymmetric deformations. If such modes exist, and are accessible in cell membranes, their being triggered would naturally lead to a reduction in the energy barrier to constriction and scission. Building on this conventional understanding of buckling analysis of thin-walled structures, we predict the existence of lower energy modes of constriction in membrane tubules. The conclusions from our simulations provide insight to a number of recent experimental studies and suggest new experimental design as discussed below. 

 Dynamin and dynamin-related proteins (DRPs) have been shown to be essential for scission events during mitochondrial division \cite{ingerman2005dnm1} and during clathrin-mediated endocytosis via mechanical feedback with actin in both yeast \cite{ fujimoto2010arabidopsis, palmer2015dynamin} and mammalian cells \cite{taylor2012feedback}. In dynamin-mediated fission during endocytosis \cite{conner2003regulated, pucadyil2008real}, dynamin preferentially interacts with curved membranes \cite{roux2010membrane, ramachandran2008real}, indicating a curvature dependence. Morlot \textit{et al.} \cite{morlot2012membrane} showed that the local energy barrier to constriction is lower at the edge of the dynamin helix (large curvature) in optical tweezer experiments of dynamin-mediated fission. More recent experiments and models suggest that fission can also occur in the middle of the dynamin-coated region \cite{dar2015high, martina2018role}. While our results cannot confirm where fission will occur, we predict two important effects - (1) constriction is indeed curvature-dependent and (2) the membrane shape at the center of a given helical pitch is highly curved in 3D. These predictions are consistent with observations \cite{Kukulski2012,morlot2012membrane,dar2015high}. For example, Dar \textit{et al.} \cite{dar2015high} showed that dynamin1 polymers cause membrane constriction with high probability when the tubule radius approaches 16 nm or less, consistent with predictions from our model (see Figure 2F of \cite{dar2015high} and compare against \Cref{fig:fig_helix}E, J). 
 
 A central conclusion from this study is that a crushed soda can shape of the neck is energetically favorable for constriction over radially symmetric pinching. This prediction suggests that mechanisms such as those proposed in Figure 6 of Dar \textit{et al}. can be revised to include lower degrees of symmetry (compare Figure 6 of \cite{dar2015high} with \Cref{fig:fig_helix}B). With advances in 3D imaging methods such as electron tomography, it should be possible to examine the cross-sections of necks during the progression of constriction by different molecular machines and quantify the relationship between membrane tubule symmetry and the particular protein assembly. Furthermore, determining the curvature-dependent rate constants for these proteins binding to the membrane will be important to quantify the relationship between the shape of the buckled membrane and the disassembly of monomers from polymerizing helical filaments such as dynamin \cite{roux2010membrane, lee2015negative}. We predict that this feedback between membrane curvature and kinetics of helix assembly-disassembly is particularly important for the membrane curvatures where our simulations determine that it is energetically expensive for multiple rings to achieve the progression to scission.
 
 

Our results also apply to cases where dynamin is not involved in the scission process. In the absence of dynamin, BAR domain proteins and actin are thought to work closely in the formation of long tubular necks \cite{ferguson2009coordinated}. Indeed, in dynamin and clathrin-independent endocytosis, actin is the primary driver of scission of tubular invaginations via a constriction force \cite{romer2010actin}. These observations suggest that while scission may be less efficient, it is still functional in the absence of dynamin. Our results show that cylindrical collars, such as those enforced by actin, can promote constriction in the absence of dynamin.

From a structural mechanics standpoint, the differences in the membrane responses to helical versus cylindrical collars can also be understood by drawing analogies again with the buckling of thin cylindrical tubes. The distributed radial pressure in a helical collar creates an lateral torque that induces a bending moment on the tube. Under these conditions, the cylindrical tubes are now susceptible to both radial collapse (through pinching) and buckling under bending moment, which can cause accelerated pinching. The soda can shape also locally reduces the area moment of inertia and this can induce buckling through a process called buckling by ovalization \cite{taylor2012shape}. In a completely different setting, the helical structures of twining vines are also known to exert squeezing forces on the support rods, suggesting that helical structures as force generating mechanisms are quite common in nature at different scales \cite{isnard2009tensioning}.  The helical collar mechanism opens up a wider parameter space (helical pitch, collar height, lateral bending, squeeze induced by helical twist, etc.) to optimize for achieving effective pinching. Such analogies with common engineering principles and with biological materials can help build our intuition on membrane-protein interactions; however, we note that the results presented in this work are specific to elastic, incompressible membranes only. 


Based on the insights derived from our simulations, future work should include further complexities such as the influence of the structure of the helical polymer, the compositional heterogeneity of cellular membranes and the effect of contact constraints between the protein and tubule that can permit potential sliding of the protein on the tubule during the scission process. While recent molecular dynamics (MD) simulations of dynamin-mediated fission also reveal non axisymmetric pathways of constriction via the formation of transient pores \cite{martina2018role}, better connections between continuum descriptions of the lipid bilayer and membrane-protein interactions at the mesoscale need to be developed to close this gap. This is an ongoing research effort in our group. 

\section*{Acknowledgements}
We would like to thank David Drubin, Jasmine Nirody, and Morgan Chabanon for their feedback on the study. P.R. would like to acknowledge the Office of Naval Research N00014-17-1-2628. S.R. would like to acknowledge the Wisconsin Alumni Research Foundation (WARF) and the Grainger Institute for Engineering at UW-Madison for funding support, and thank Prof. Xiaoping Qian at UW-Madison for his advise on geometric modeling. M.A. would like to acknowledge the Arnold O. Beckman Postdoctoral Fellowship. KG would like to acknowledge NSF DMREF grant \#1729166.
\newpage
\bibliography{ref}

\newpage

\beginsupplement
\section*{Supplementary Online Material}

\section*{Assumptions} 

\begin{itemize}

\item The lipid bilayer is a modeled as a thin elastic shell. We use the Helfrich energy \cite{helfrich1973elastic} based on the assumption that the thickness of the bilayer is negligible compared to its radius of curvature. This allows us to neglect shear deformations and consider classical Kirchoff-Love shell kinematics for thin shell geometries. Furthermore, we assume that the membrane is areally incompressible since the maximum elastic stretch is only 4 $\%$ \cite{evans1979mechanics}. This incompressibility constraint is numerically enforced using a Lagrange multiplier field. Additionally, we ignore any fluid \cite{rangamani2013interaction} and friction \cite{simunovic2017friction} properties of the bilayer. Thus, the membrane is in mechanical equilibrium at all times. 
\item The lack of resistance to shear deformation modes in the Helfrich energy formulation can result in rigid body (zero energy) modes of deformation. To circumvent this limitation, in the 3D numerical simulations, we add shear stabilization terms to the classical Helfrich energy functional \cite{sauer2017}. These stabilization terms are of a smaller magnitude relative to the traditional bending energy terms, and restore stability to the numerical model without significantly effecting the kinematics of bending. 
\item The membrane tubule is modeled both as an axisymmetric and 3D lipid bilayer. A pinching force is applied at different locations - `cap', `tube', and `base' (Fig. \ref{fig:schematic}). Since the tether is pulled from a membrane reservoir that can buffer changes in membrane tension 
\cite{raucher1999characteristics, mulholland1994ultrastructure}, we assume that elastic properties like membrane tension and bending rigidity are constant. 
\item  Since we do not consider the fluid properties of the membrane, we cannot consider scission explicitly. We assume that the large stresses at the neck can lead to the formation of a hemi-fission intermediate \cite{walani2015endocytic,kozlov2001fission}.
\item The interaction of the constriction proteins and the membrane tubule can be numerically thought of as a contact model where the proteins apply a contact force of constriction on the tubule. Here, we do not consider a contact model but rather apply a follower load type collar pressure in the constriction region.     
\end{itemize}

\section*{Thin shell formulations: Axisymmetric and 3D models}

\subsection*{Equilibrium equations for the axisymmetric model}\label{sec:s-equilibrium}
First, we write the force balance on the membrane as

\begin{align} 
\nabla \cdot \boldsymbol{\sigma}+p\textbf{n}=\textbf{f}, 
\label{eq:s-Newton}
\end{align}

\noindent where $\boldsymbol{\sigma}$ is the stress tensor, p is the pressure difference between the inside and outside of the volume bounded by the membrane, and $\textbf{f}$ is any externally applied force per unit area on the membrane. In our simulations, we assume that the tubule has equilibrated the pressure difference, and thus set p to 0. $\textbf{f}$ includes both the axial and pinching forces applied on the membrane. By introducing the covariant derivative as $()_{;\alpha}$, the surface divergence in Eq. \ref{eq:s-Newton} can be rewritten as \citep{steigmann1999fluid}

 \begin{align} 
\nabla \cdot \boldsymbol{\sigma}= \boldsymbol{\sigma}^{\alpha}_{;\alpha}=(\sqrt{a})^{-1} (\sqrt{a}\boldsymbol{\sigma}^{\alpha})_{,\alpha},
\label{eq:s-divergence}
\end{align}

\noindent where $a$ is the determinant of the first fundamental form metric $a_{\alpha \beta}$. The surface stresses in Eq. \ref{eq:s-Newton} can be split into normal and tangential component given by

 \begin{align} 
\boldsymbol{\sigma}^{\alpha}= \boldsymbol{T}^{\alpha}+S^{\alpha} \textbf{n},
\label{eq:s-stress decomposition}
\end{align}

\noindent where

\begin{align} 
\textbf{T}^{\alpha} =T^{\alpha \beta} \textbf{a}_{\beta}, \quad \quad T^{\alpha \beta}=\sigma^{\alpha \beta}+b^{\beta}_{\mu}M^{\mu \alpha}, \quad  \quad S^{\alpha}=-M^{\alpha \beta}_{; \beta}.
\label{eq:s-stress1}
\end{align}

The two tensors $\sigma^{\alpha \beta}$ and $M^{\alpha \beta}$ can be expressed by the derivative of $F$, the energy per unit mass, with respect to the coefficients of the first and second fundamental forms,  $a_{\alpha \beta}$, $b_{\alpha \beta}$, respectively \cite{rangamani2013interaction,steigmann1999fluid}

\begin{align} 
\sigma^{\alpha \beta}=\rho (\frac{\partial F(\rho,H,K;x^{\alpha})}{\partial a_{\alpha\beta}}+\frac{\partial F(\rho,H,K;x^{\alpha})}{\partial a_{\beta\alpha}}), \\ M^{\alpha\beta}=\frac{\rho}{2} (\frac{\partial F(\rho,H,K;x^{\alpha})}{\partial b_{\alpha\beta}}+\frac{\partial F(\rho,H,K;x^{\alpha})}{\partial b_{\beta\alpha}}),
\label{eq:s-F derivative}
\end{align}

\noindent where $\rho$ is the surface mass density. $H$ and $K$ are mean and Gaussian  curvatures  given by

\begin{align} 
H=\frac{1}{2} a^{\alpha \beta}b_{\alpha \beta}, \quad K=\frac{1}{2} \varepsilon^{\alpha \beta} \varepsilon ^{\lambda \mu} b_{\alpha \lambda}b_{\beta \mu}.
\label{eq:s-Curvatures}
\end{align}

Here $(a^{\alpha\beta})=(a_{\alpha \beta})$ is the dual metric and $\varepsilon ^{\alpha \beta}$ is the permutation tensor defined by $\varepsilon ^{12}=-\varepsilon ^{21}=\frac{1}{\sqrt{a}}, \varepsilon ^{11}=\varepsilon ^{22}=0$. 

A reasonable assumption to make is that the membrane tubule has a fixed area. We introduce an area incompressibility ($J=1$) constraint using a general form of free energy density per unit mass given as

\begin{align} 
F(\rho,H,K;x^{\alpha})=\tilde{F}(H,K;x^{\alpha})-\frac{\gamma (x^{\alpha},t)}{\rho}.
\label{eq:s-Lagrange multiplier}
\end{align}

Here $\gamma (x^{\alpha}, t)$ is a Lagrange multiplier field required to impose invariance of $\rho$ on the whole of the surface (see \cite{steigmann1999fluid} for full derivation). Substituting $W=\rho \tilde{F}$ into Eq. \ref{eq:s-Lagrange multiplier} we get

\begin{align} 
\sigma^{\alpha \beta}=(\lambda +W)a^{\alpha \beta} -(2HW_H+2\kappa W_K)a^{\alpha \beta} +W_H\tilde{b}^{\alpha \beta}, \\
M^{\alpha \beta}=\frac{1}{2}W_H a^{\alpha \beta} +W_K \tilde{b}^{\alpha \beta},
\label{eq:s-stress2}
\end{align}

\noindent where 

\begin{align} 
\lambda=-(\gamma+W).
\label{eq:s-lambdaa}
\end{align}

Combining Eqs. \ref{eq:s-stress2}, \ref{eq:s-stress1}, and \ref{eq:s-stress decomposition} into Eq. \ref{eq:s-Newton} gives the equations in normal and tangential equations as

\begin{align} 
p+\textbf{f} \cdot \textbf{n}=\Delta{\frac{1}{2} W_H}+(W_K)_{; \alpha \beta} \tilde{b}^{\alpha \beta} +W_H(2H^2- K) \nonumber  \\ 
+2H(KW_K-W)-2 \lambda H,
\label{eq:s-normal}
\end{align}

and

\begin{align} 
N^{\beta \alpha}_{; \alpha}-S^{\alpha}b^{\beta}_{\alpha}=-(\gamma_{, \alpha} +W_K k_{, \alpha}+W_H H_{, \alpha}) a^{\beta \alpha} \nonumber \\ 
= (\frac{\partial{W}}{\partial{x^{\alpha}_{| exp}}}+\lambda_{, \alpha}) a^ {\beta \alpha} = \textbf{f}\cdot\textbf{a}_s.
\label{eq:s-Tangential}
\end{align}
 
Here $\Delta (\cdot)$ is the surface Laplacian and $()_{| exp}$ denotes the explicit derivative respect to coordinate $\theta^{\alpha}$.

\subsection*{Axisymmetric model}
Using the axisymmetric parametrization
 
\begin{equation}
\textbf{r}(s, \theta) = r(s)\textbf{e}_{r}(\theta) + z(s)\textbf{k}.
\end{equation} 

 we define $\psi$ as the angle made by the tangent with respect to the horizontal (see Fig. \ref{fig:coordinates}). This gives $r'(s) =  \cos(\psi)$, $z'(s) =  \sin(\psi)$, which satisfies the identity $(r')^2+ (z')^2 = 1$. Using this, we  define the normal to the surface as $\textbf{n}=-\sin\psi\textbf{e}_{r}(\theta)+\cos\psi\textbf{k}$, the tangent to the surface in the direction of increasing arc length as $\textbf{a}_{s}=\cos\psi\textbf{e}_{r}(\theta)+\sin\psi\textbf{k}$, and unit vector \textbf{$\tau$} = $\textbf{e}_{\theta}$ tangent to the boundary $\partial \omega$ in the direction of the surface of revolution. For more details, we refer the reader to \cite{steigmann1999fluid,hassinger2017design,rangamani2014protein}.

The expressions for tangential $(\kappa_{\nu})$, transverse $(\kappa_{\tau})$ and twist $(\tau)$ curvatures are simplified as

\begin{equation}
\kappa_{\nu}=\psi^{'},\quad \kappa_{\tau}=r^{-1}\sin\psi,\quad \tau=0.
\label{eq:s-kappas}
\end{equation}

Further, we calculate the mean curvature ($H$) and Gaussian curvature ($K$) as

\begin{equation}
H=\frac{1}{2}(\kappa_\nu+\kappa_\tau)=\frac{1}{2}(\psi^{'}+r^{-1}\sin\psi),  \quad
K=\kappa_{\tau}\kappa_{\nu}=\frac{\psi^{'}\sin\psi}{r}.
\label{eq:s-curvatures}
\end{equation}

We introduce a term $L=\frac{1}{2\kappa}r(W_H)'$ in order to write a system of first order differential equations governing the problem \cite{hassinger2017design} as ,

\begin{align}   
{r}' &= \cos{\psi}, \quad  {z}' = \sin{\psi},  \nonumber \\ r{\psi}' &= 2 r H - \sin{\psi}, \quad r {H}' = L + r {C}', \nonumber \\ \frac{L'}{r} &= \frac{p}{k} + \frac{\mathbf{f} \cdot \mathbf{n}}{\kappa} + 2H \left[ \left(H - C \right)^2 + \frac{\lambda}{\kappa} \right] \nonumber \\ &- 2 \left( H - C \right) \left[ H^2 + \left( H - r^{-1} \sin{\psi} \right)^2 \right] \nonumber, \\ {\lambda}' &= 2 {\kappa} \left( H - C \right) {C}' -  \mathbf{f} \cdot \mathbf{a_s}.
\label{eq:s-systemofequations}
\end{align}

Eq. \ref{eq:s-systemofequations} is a function of the arc length (s). This can be rewritten in terms of membrane area (a) using

\begin{equation}
a(s)=2\pi \int_0^s r(\xi)d \xi\quad \rightarrow \quad \frac{da}{ds}=2\pi r.
\label{eq:s-area-arclength}
\end{equation}

We choose to non-dimensionalize our system of equations using a length scale $R_{0}$ and bending rigidity scale $\kappa_{0}$ as

\begin{align}
\begin{split}
\alpha=\frac{a}{2 \pi R_0^2},\quad x=\frac{r}{R_0}, \quad y=\frac{y}{R_0}, \quad h=HR_0, \quad c=CR_0, \quad l=LR_0,\\ \lambda^{*}=\frac{\lambda R_0^2}{\kappa_0}, \quad p^*=\frac{pR_0^3}{\kappa_0}, \quad f^*=\frac{fR_0^3}{\kappa_0}, \quad \kappa^*=\frac{\kappa}{\kappa_0},
\label{eq:s-non-dimension}
\end{split}
\end{align}

where R$_{0}$ is the radius of the flat patch of membrane in simulations of a membrane tubule (Fig. \ref{fig:schematic}E), the radius of the hemisphere for simulations of the tubule cap (Fig. \ref{fig:schematic}B), and the radius of the tube for simulations of the tube (Fig. \ref{fig:schematic}C) and the base (Fig. \ref{fig:schematic}D). $\kappa_{0}$ is the bending rigidity of the bare membrane. 

Rewriting Eq. \ref{eq:s-systemofequations} using the dimensionless variables in Eq. \ref{eq:s-non-dimension}, we get \cite{hassinger2017design}

\begin{align}
x\dot{x} &= \cos{\psi}, \quad  x\dot{y} = \sin{\psi}, \nonumber \\ \quad x^2\dot{\psi} &= 2 x h - \sin{\psi}, \quad x^2 \dot{h} = l + x^2 \dot{c}, \nonumber \\ \dot{l} &= \frac{p^*}{\kappa^*} + \frac{\mathbf{f}^* \cdot \mathbf{n}}{\kappa^*} + 2h \left[ \left(h - c \right)^2 + \frac{\lambda^*}{\kappa^*} \right] \nonumber \\ &- 2 \left( h - c \right) \left[ h^2 + \left( h - x^{-1} \sin{\psi} \right)^2 \right], \nonumber \\ \dot{\lambda^*} &= 2 \kappa^* \left( h - c \right) \dot{c} -  \frac{\mathbf{f}^* \cdot \mathbf{a_s}}{x}.
\label{eq:s-area-systemofequations}
\end{align}

We enforce a third boundary point for constriction simulations at the base (Fig. \ref{fig:schematic}D) and the whole tube (Fig. \ref{fig:schematic}E) by introducing an independent variable \cite{shampine2000solving}

\begin{equation}
\zeta = \alpha_{bp} \frac{\alpha - \alpha_{bp}}{\alpha_{tot} - \alpha_{bp}}, 
\end{equation}

where $\alpha_{bp}$ is the non-dimensional area of the first `phase' and $\alpha_{tot}$ is the total non-dimensional area of the membrane. $\alpha $ is the variable defining the non-dimensional area along the first `phase' and $\zeta$ is the variable defining non-dimensional area along the second `phase'. Like $\alpha$ in the first interval, $\zeta$ ranges from o to $\alpha_{bp}$ in the second interval. Thus we can redefine our system of equations (Eq. \ref{eq:s-area-systemofequations}) for 2 phases as

\begin{align}
x_{1}\frac{dx_{1}}{d\alpha} &= \cos{\psi_{1}}, \quad  x_{1}\frac{dy_{1}}{d\alpha} = \sin{\psi_{1}}, \nonumber \\ \quad x_{1}^2\frac{d\psi_{1}}{d\alpha} &= 2 x_{1} h_{1} - \sin{\psi_{1}}, \quad x_{1}^2 \frac{dh_{1}}{d\alpha} = l_{1} + x_{1}^2 \dot{c_{1}}, \nonumber \\ \frac{dl_{1}}{d\alpha} &= \frac{p^*}{\kappa^*} + \frac{\mathbf{f_{1}^*} \cdot \mathbf{n}}{\kappa^*} + 2h_{1} \left[ \left(h_{1} - c \right)^2 + \frac{\lambda_{1}^*}{\kappa^*} \right] \nonumber \\ &- 2 \left( h_{1} - c_{1} \right) \left[ h_{1}^2 + \left( h_{1} - x_{1}^{-1} \sin{\psi_{1}} \right)^2 \right], \nonumber \\ \frac{d\lambda_{1}^*}{d\alpha} &= 2 \kappa^* \left( h_{1} - c_{1} \right) \dot{c_{1}} -  \frac{\mathbf{f_{1}^*} \cdot \mathbf{a_s}}{x} ,\nonumber \\
x_{2}\frac{dx_{2}}{d\zeta} &= (\frac{\alpha_{tot} - \alpha_{bp}}{\alpha_{bp}})\cos{\psi_{2}}, \quad  x_{2}\frac{dy_{2}}{d\zeta} = (\frac{\alpha_{tot} - \alpha_{bp}}{\alpha_{bp}})\sin{\psi_{2}}, \nonumber \\ \quad x_{2}^2\frac{d\psi_{2}}{d\alpha} &= (\frac{\alpha_{tot} - \alpha_{bp}}{\alpha_{bp}}) (2 x_{2} h_{2} - \sin{\psi_{2}}), \nonumber 
\\ x_{2}^2 \frac{dh_{2}}{d\zeta} &= (\frac{\alpha_{tot} - \alpha_{bp}}{\alpha_{bp}})(l_{2} + x_{2}^2) \dot{c_{2}}, \nonumber \\ \frac{dl_{2}}{d\zeta} &= (\frac{\alpha_{tot} - \alpha_{bp}}{\alpha_{bp}})(\frac{p^*}{\kappa^*} + \frac{\mathbf{f_{2}^*} \cdot \mathbf{n}}{\kappa^*} + 2h_{2} \left[ \left(h_{2} - c_{2} \right)^2 + \frac{\lambda_{2}^*}{\kappa^*} \right] \nonumber \\ &- 2 \left( h_{2} - c_{2} \right) \left[ h_{2}^2 + \left( h_{2} - x_{2}^{-1} \sin{\psi_{2}} \right)^2 \right]), \nonumber \\ \frac{d\lambda_{2}^*}{d\zeta} &= (\frac{\alpha_{tot} - \alpha_{bp}}{\alpha_{bp}})(2 \kappa^* \left( h_{2} - c_{2} \right) \dot{c_{2}} -  \frac{\mathbf{f_{2}^*} \cdot \mathbf{a_s}}{x}),
\label{eq:s-area-systemofequations2}
\end{align}

where 1 and 2 are the two phases represented by non-dimensional areas $\alpha$ and $\zeta$, both of which are defined in the interval [0 $\alpha_{bp}$]. Applied forces are modeled as a smooth hyperbolic tangent function. For example, the axial force at the tip of the tubule and radial force at the interface are modeled as

\begin{align}
f_{axial} = f_{ax} \times (0.5(1 - \tanh(\text{g}*(\mathbf{\alpha} - \alpha_{\text{axial}}))))/\alpha_{\text{axial}}, \\
f_{pinch} = f_{p} \times 0.5(\tanh(g(\mathbf{\alpha} - (\alpha_{bp} - \alpha_{\text{radial}}))) \nonumber \\ - 0.5\tanh(g(\mathbf{\alpha} - \alpha_{bp}))),
\label{eq:s-C}
\end{align}

where $f_{axial}$ and $f_{pinch}$ are the distribution of axial and radial forces per unit area along the non-dimensional membrane area $\alpha$, $f_{ax}$ and $f_{p}$ are the magnitudes for force per unit area, and $\alpha_{\text{axial}}$ and $\alpha_{\text{radial}}$ are the corresponding non-dimensional areas of applied force in the axial and radial direction respectively. g is a constant that ensures a sharp but smooth transition. In our simulations, we use g = 20 \cite{hassinger2017design}. 




\subsubsection*{Boundary conditions}

Eq \ref{eq:s-area-systemofequations2} can be solved given a set of boundary conditions. All the axisymmetric simulations in this study were performed using the MATLAB bvp4c toolbox \cite{shampine2000solving}. A useful feature of this toolbox is the estimation of unknown parameters by providing additional boundary conditions \cite{shampine2000solving}. The MATLAB subroutines used in this work are available on Github \cite{Vasan2019}.

\begin{itemize}
\item{ \textbf{Whole tubule}}

The tubule pinching simulations in Fig. \ref{fig:schematic}E were performed using the following set of boundary conditions

\begin{align}
x_{1}(0) = 0, \quad x_{1}(\alpha_{bp}) = \mathbf{x_{p}}, \quad \psi_{1}(0) = 0, \quad \psi_{2}(\alpha_{bp}) = 0,\nonumber \\
y_{1}(0) = \mathbf{y_{p}}, \quad y_{2}(\alpha_{bp}) = 0, \quad , \lambda_{2}(\alpha_{bp}) = \mathbf{\lambda_{0}}, \quad l_{1}(0) = 0, \nonumber\\
x_{1}(\alpha_{bp}) = x_{2}(0), \quad y_{1}(\alpha_{bp}) = y_{2}(0), \quad \psi_{1}(\alpha_{bp}) = \psi_{2}(0), \nonumber \\ h_{1}(\alpha_{bp}) = h_{2}(0), \quad l_{1}(\alpha_{bp}) = l_{2}(0), \quad \lambda_{1}(\alpha_{bp}) = \lambda_{2}(0).
\label{BC_tubule}
\end{align}

$\mathbf{x_{p}}$ and $\mathbf{y_{p}}$ are additional constraints for the radius at the interface and height of the tubule respectively. These additional constraints are used to estimate the axial force and pinching force required to obtain a solution to the system of equations in Eq. \ref{eq:s-area-systemofequations2}. $\mathbf{\lambda_{0}}$ is the boundary membrane tension. We note here that this now represents a system of 12 equations and 2 unknown parameters with 14 boundary conditions. The 2 unknown parameters are the axial and radial force. 6 of these boundary conditions are continuity conditions for every parameter at the interface (Eq. \ref{BC_tubule}).

\item{\textbf{Base}}

The half catenoid-like base pinching simulations in Fig. \ref{fig:schematic}D were performed using the following set of boundary conditions

\begin{align}
x_{1}(0) &= \mathbf{x_{0}}, \quad x_{1}(\alpha_{bp}) = \mathbf{x_{p}}, \quad \psi_{1}(0) = \pi, \quad \psi_{2}(\alpha_{bp}) = \pi/2, \nonumber \\
y_{1}(0) &= 0,  \quad , \lambda_{1}(0) = \mathbf{\lambda_{0}}, \quad l_{2}(\alpha_{bp}) = 0, \nonumber \\
x_{1}(\alpha_{bp}) &= x_{2}(0), \quad y_{1}(\alpha_{bp}) = y_{2}(0), \quad \psi_{1}(\alpha_{bp}) = \psi_{2}(0), \nonumber \\ h_{1}(\alpha_{bp}) &= h_{2}(0), \quad l_{1}(\alpha_{bp}) = l_{2}(0), \quad \lambda_{1}(\alpha_{bp}) = \lambda_{2}(0).
\end{align}

$\mathbf{x_{p}}$ is the additional constraint for the radius at the interface. $\mathbf{x_{0}}$ is the radius at y = 0. The additional constraint is used to estimate the pinching force required to obtain a solution to the system of equations in Eq. \ref{eq:s-area-systemofequations2}. $\mathbf{\lambda_{0}}$ is the boundary membrane tension. We note here that this now represents a system of 12 equations and 1 unknown parameter with 13 boundary conditions. 

\item{\textbf{Tube}}

The cylinder/tube pinching simulations in Fig. \ref{fig:schematic}C were performed by solving the system of equations in Eq. \ref{eq:s-area-systemofequations} for a single phase of membrane. The boundary conditions used were

\begin{align}
l(0) &= 0, \quad y(0) = 0, \quad y(\alpha) = Z_{0}/R_{0}, \quad \psi(0) = \pi/2, \nonumber \\
x(\alpha) &= 1, \quad x(0) = \mathbf{x_{p}}, \quad \psi{(\alpha)} = \pi/2,
\end{align}

where $\alpha$ is the non-dimensional area of the tube, $\mathbf{x_{p}}$ is the additional constraint required for estimating the pinching force. This represents a system of 6 equations and 1 unknown parameter with 7 boundary conditions. 

\item{\textbf{Cap}}

The hemisphere/cap pinching simulations in figure \ref{fig:schematic}B were performed by solving the system of equations in Eq. \ref{eq:s-area-systemofequations} for a single phase of membrane. The boundary conditions used were
 
\begin{align}
l(0) &= 0, \quad y(0) = 0, \quad l(\alpha) = 0, \quad \psi(0) = \pi/2, \nonumber \\
x(\alpha) &= 0, \quad x(0) = \mathbf{x_{p}}, \quad \psi{(\alpha)} = \pi,
\end{align}

where $\mathbf{x_{p}}$ is the additional constraint required for estimating the pinching force. This represents a system of 6 equations and 1 unknown parameter with 7 boundary conditions. 
\end{itemize}

\subsection*{3D model}
\subsubsection*{Thin shell formulation}
Considering the classical Helfrich formulation for biological membranes, the strain energy density of a membrane in the current configuration is given by:
\begin{equation}
W = k_B (H-H_0)^2+k_G \kappa
\end{equation}
where $k_B$ and $k_G$ are the bending modulus and the Gaussian modulus of the membrane, $H$ is the mean curvature, $\kappa$ is the Gaussian curvature and $H_0$ represents the instantaneous curvature induced in the membrane.  

To enforce area-incompressibility, we consider the following Lagrange multiplier formulation:
\begin{equation}
W_{LM} = J(k_B (H-H_0)^2+k_G \kappa) + q (J-1)
\end{equation}
where q is the point value of the Lagrange multiplier field, and J is the surface stretch (ratio of area in the current configuration to the area in the reference configuration).  

The governing equation for quasi-static mechanical equilibrium in 3D simulations is obtained by minimizing the Helfrich energy functional following standard variational arguments, and is given by \cite{sauer2017}:  
\begin{equation}
\int_{\partial \Omega} \frac{1}{2} \delta a_{ij} \sigma^{ij} ~da + \int_{\partial \Omega} \delta b_{ij} M^{ij} ~da - \int_{\partial \Omega_{\text{collar}}} \delta \boldsymbol{x} \cdot \boldsymbol{p} ~da - \int_{\Gamma} \delta \boldsymbol{x} \cdot \boldsymbol{t} ~ds = 0
\label{EqnWeakForm}
\end{equation}
where $\partial \Omega$ is the membrane surface and $\Gamma$ is the membrane boundary on which surface tractions can be applied, as shown in \Cref{fig:coordinates}B. $\delta a_{ij}$ and $\delta b_{ij}$ are variations of the components of the metric tensor and the curvature tensor, respectively. $\sigma^{ij}$ are the components of the stress tensor, $M^{ij}$ are components of the moment tensor, $\boldsymbol{p}$ is the pressure applied along a collar on the membrane surface (to cause constriction), and $\boldsymbol{t}$ is the surface traction. 

For a hyperelastic material model, we can express the stress and moment components in terms of the strain energy density as \cite{Duong2017}: 
\begin{eqnarray}
\sigma^{ij} = \frac{2}{J} \frac{\partial W}{\partial a_{ij}}\\
M^{ij} = \frac{1}{J} \frac{\partial W}{\partial b_{ij}}
\end{eqnarray}
For the Helfrich type strain energy density, these take the form:
\begin{align}
\sigma^{ij} &= (k_B (H-H_0)^2 - k_G \kappa) a ^{ij} - 2 k_B (H-H_0) b^{ij}\\
M^{ij} &= (k_B (H-H_0) + 2 k_G H)a ^{ij} - k_G b^{ij}\
\end{align}

\subsubsection*{Computational implementation}
We solve the governing equation given by Eq.\ref{EqnWeakForm} using a Isogeometric Analysis (IGA) based numerical framework for solving problems of membrane mechanics developed as part of this work. A companion manuscript (in preparation by the authors) describes the details of the mathematical methods and the numerical formulation. The computational implementation, along with the source code for solving the boundary value problems listed below, is available as a public code repository on GitHub \cite{Rudraraju2019}. 

\subsubsection*{List of 3D simulations}
For each of the 3D simulations, we solve the governing equation given by Eq.\ref{EqnWeakForm} using a force control or displacement control approach, with the relevant displacement, angle and traction boundary conditions. The displacement boundary conditions are applied on the components of the displacement vector, $\bu$, that is defined as the change in position of a point on the membrane between its current and reference configuration ($\bu(\xi_1,\xi_2) = \textbf{x}(\xi_1,\xi_2)-\textbf{X}(\xi_1,\xi_2)$). The angle boundary conditions, where needed, are enforced through the weak formulation using a penalty approach and result in the normal vector ($\textbf{n}$) at the boundaries to align along the preferred direction. In all the simulations, we have two boundaries, and these are identified as the inner boundary ($\Gamma_{\text{I}}$) and the outer boundary ($\Gamma_{\text{O}}$) as indicated in the schematic in \Cref{fig:coordinates}B. The specific numerical simulations in this work using the 3D model are described below. 

\begin{itemize}
\item{\textbf{Tube pulling}}

The tube pulling simulation shown in \Cref{fig:pullout_comparison}(A) considers a reference circular plate geometry with an outer radius of 20 nm, and an inner radius of 0.2 nm. The boundary value problem is solved as a force control problem with a traction on the inner boundary ($\Gamma_{\text{I}}$). The displacement and traction boundary conditions are as follows:
\begin{align*}
t_{y}  &= h \quad \mathrm{on} \quad \Gamma_{\text{I}} \\
u_{x}  &= 0 \quad \& \quad u_{z}  = 0 \quad \mathrm{on} \quad \Gamma_{\text{I}} \\
u_{y}  &= 0 \quad \mathrm{on} \quad \Gamma_{\text{O}}
\end{align*}
See Movie M12 in the supplementary information for the evolution of the membrane deformation.

\item{\textbf{Whole tubule}}

The whole tubule simulations shown in \Cref{fig:fig_3D_whole_tube} consider pinching at three different locations, identified as the cap, tube and base locations (\Cref{fig:schematic}A). For the tubule geometry, shown in \Cref{fig:schematic}A, the tubule radius is 20 nm and height is 100 nm. The inner boundary ($\Gamma_{\text{I}}$) at the top of the tubule has a radius of 0.2 nm and the outer boundary ($\Gamma_{\text{O}}$) at the base of the tubule has a radius of 40 nm. The boundary value problem is solved as a force control problem with pressure applied on a collar ($\Omega_{\text{collar}}$) located at the cap, tube or base location. In addition, the displacement boundary conditions are as follows:
\begin{align*}
u_{x}  &= 0 \quad \mathrm{on} \quad \Gamma_{\text{I}} \\
u_{y}  &= 0 \quad \mathrm{on} \quad \Gamma_{\text{I}} \\
u_{z}  &= 0 \quad \mathrm{on} \quad \Gamma_{\text{I}} 
\end{align*}
See Movies M1-M3 in the supplementary information for the evolution of the constriction process for the cap, tube and base locations.

\item{\textbf{Base}}

The constriction simulation shown in \Cref{fig:comparison_base_pinching} considers pinching at the base location. The tube geometry considered has a radius of 20 nm, and a height of 80 nm. The tube boundary on the top is identified as the inner boundary ($\Gamma_{\text{I}}$) and the tube boundary at the bottom is identified as the outer boundary ($\Gamma_{\text{O}}$). The boundary value problem is solved as a force control problem with pressure applied on a collar ($\Omega_{\text{collar}}$) located at the base location. In addition, the displacement boundary conditions are as follows:
\begin{align*}
u_{x}  &= 0 \quad \mathrm{on} \quad \Gamma_{\text{I}} \\
u_{y}  &= 0 \quad \mathrm{on} \quad \Gamma_{\text{I}} \\
u_{z}  &= 0 \quad \mathrm{on} \quad \Gamma_{\text{I}} 
\end{align*}
See Movie M3 in the supplementary information for the evolution of the constriction process for the base location.

\item{\textbf{Cap}}

The constriction simulation shown in \Cref{fig:disp_control_cap_pinching} considers pinching at the cap location. The cap geometry is a hemisphere with a radius of 20 nm. The cap boundary on the top, with a small radius of 1 nm, is identified as the inner boundary ($\Gamma_{\text{I}}$) and the cap boundary at the bottom is identified as the outer boundary ($\Gamma_{\text{O}}$). This boundary value problem is solved as a displacement control problem, as the force control problem is numerically unstable due to the rigid body modes induced under the displacement boundary conditions considered. As this problem is solved as a displacement control problem, this enforces axisymmetry of the pinching profile. The displacement boundary conditions are as follows:
\begin{align*}
u_{x}  &= g \quad \mathrm{on} \quad \Gamma_{\text{O}} \\
u_{y}  &= 0 \quad \mathrm{on} \quad \Gamma_{\text{O}} \\
u_{z}  &= g \quad \mathrm{on} \quad \Gamma_{\text{O}} 
\end{align*}
See Movie M13 in the supplementary information for the evolution of the constriction process for the cap location.

\item{\textbf{Tube}}

The constriction simulation shown in \Cref{fig:disp_control_tube_pinching} considers pinching at the tube location. The tube geometry is a cylinder with a radius of 20 nm. The tube boundary on the top is identified as the inner boundary ($\Gamma_{\text{I}}$) and the boundary at the bottom is identified as the outer boundary ($\Gamma_{\text{O}}$). This boundary value problem is solved as a displacement control problem. Like in the case of the cap simulation, as this problem is solved as a displacement control problem, this enforces axisymmetry of the pinching profile. The displacement boundary conditions are as follows:
\begin{align*}
u_{x}  &= g \quad \mathrm{on} \quad \Gamma_{\text{O}} \\
u_{z}  &= g \quad \mathrm{on} \quad \Gamma_{\text{O}} \\
u_{y}  &= 0 \quad \mathrm{on} \quad \Gamma_{\text{I}} 
\end{align*}
See Movie M14 in the supplementary information for the evolution of the constriction process with displacement control for the tube location.\\

We also solve a force control equivalent of this problem, and this is shown in \Cref{fig:force_control_tube_pinching}. This case is discussed below in the simulation of the helical force collar at the tube location with a zero helical pitch. See Movie M15 in the supplementary information for the evolution of the constriction process with force control for the tube location.

\item{\textbf{Helical force collar at the tube location}}

The constriction simulation shown in \Cref{fig:fig_helix} considers pinching at the tube location due to helical collar. In \Cref{fig:fig_helix}A we consider a single helical ring, and in \Cref{fig:fig_helix}G we consider three helical rings.  The tube geometry considered for the single helical ring case has a radius of 20 nm, and a height of 40 nm. The tube geometry considered for the three helical rings case has a radius of 20 nm, and a height of 200 nm. For both cases, the tube boundary on the top is identified as the inner boundary ($\Gamma_{\text{I}}$) and the tube boundary at the bottom is identified as the outer boundary ($\Gamma_{\text{O}}$). The boundary value problem is solved as a force control problem with pressure applied on a helical collar ($\Omega_{\text{collar}}$) located at the tube location. In addition, the displacement boundary conditions are as follows:
\begin{align*}
u_{x}  &= 0 \quad \mathrm{on} \quad \Gamma_{\text{I}} \\
u_{y}  &= 0 \quad \mathrm{on} \quad \Gamma_{\text{I}} \\
u_{z}  &= 0 \quad \mathrm{on} \quad \Gamma_{\text{I}} \\
u_{x}  &= 0 \quad \mathrm{on} \quad \Gamma_{\text{O}} \\
u_{y}  &= 0 \quad \mathrm{on} \quad \Gamma_{\text{O}} \\
u_{z}  &= 0 \quad \mathrm{on} \quad \Gamma_{\text{O}} 
\end{align*}
See Movies M4-M6 in the supplementary information for the evolution of the constriction process due to a helical force collar at the tube location with a non-dimensional pitch of zero, two and four, respectively, and movie M7 for the corresponding evolution of the constriction process due to a force collar with three helical rings.

\item{\textbf{Helical force collar at the base location}}

The constriction simulation shown in \Cref{fig:fig_helix_base} considers pinching at the base location due to helical collar. In \Cref{fig:fig_helix_base}A we consider a single helical ring, and in \Cref{fig:fig_helix_base}G we consider three helical rings.  The tube geometry considered for the single helical ring case has a radius of 20 nm, and a height of 40 nm. The tube geometry considered for the three helical rings case has a radius of 20 nm, and a height of 200 nm. For both cases, the tube boundary on the top is identified as the inner boundary ($\Gamma_{\text{I}}$) and the tube boundary at the bottom is identified as the outer boundary ($\Gamma_{\text{O}}$). The boundary value problem is solved as a force control problem with pressure applied on a helical collar ($\Omega_{\text{collar}}$) located at the base location. In addition, the displacement boundary conditions are as follows:
\begin{align*}
u_{x}  &= 0 \quad \mathrm{on} \quad \Gamma_{\text{I}} \\
u_{y}  &= 0 \quad \mathrm{on} \quad \Gamma_{\text{I}} \\
u_{z}  &= 0 \quad \mathrm{on} \quad \Gamma_{\text{I}} 
\end{align*}
See Movies M8-M10 in the supplementary information for the evolution of the constriction process due to a helical force collar at the base location with a non-dimensional pitch of zero, two and four, respectively, and movie M11 for the corresponding evolution of the constriction process due to a force collar with three helical rings.

\end{itemize}

\section*{Analytical solution for tube pulling simulation}



The equilibrium values of R$_0$ and f$_0$ for a membrane tube are defined as

\begin{align}
    R_0 = \sqrt{\kappa/(2 \sigma)}, \\
    f_{0} = 2\pi \sqrt{(2 \sigma \kappa)},
\end{align}

where $\kappa$ is the bending rigidity, $\sigma$ is the membrane tension. For bending rigidity of 20 pN$\cdot$nm and membrane tension 0.1 pN/nm, we get

\begin{align}
    f_0 = 12.5664 \mbox{  pN},
\end{align}

which is the equilibrium value of force obtained in Fig. \ref{fig:pullout_comparison}.

\pagebreak

\subsubsection*{List of movies}
\begin{itemize}
\item Movie M1: Evolution of the constriction process for the cap location for the whole tubule geometry. 
\item Movie M2: Evolution of the constriction process for the tube location for the whole tubule geometry. 
\item Movie M3: Evolution of the constriction process for the base location for the whole tubule geometry. 
\item Movie M4: Evolution of the constriction process due to a single helical force collar at the tube location with a non-dimensional pitch of zero. 
\item Movie M5: Evolution of the constriction process due to a single helical force collar at the tube location with a non-dimensional pitch of two.
\item Movie M6: Evolution of the constriction process due to a single helical force collar at the tube location with a non-dimensional pitch of four. 
\item Movie M7: Evolution of the constriction process due to a force collar with three helical rings at the tube location. 
\item Movie M8: Evolution of the constriction process due to a single helical force collar at the base location with a non-dimensional pitch of zero. 
\item Movie M9: Evolution of the constriction process due to a single helical force collar at the base location with a non-dimensional pitch of two.
\item Movie M10: Evolution of the constriction process due to a single helical force collar at the base location with a non-dimensional pitch of four. 
\item Movie M11: Evolution of the constriction process due to a force collar with three helical rings at the base location. 
\item Movie M12: Evolution of membrane deformation for pulling of a tubule from a flat membrane.
\item Movie M13: Evolution of a axisymmetric constriction profile due to a displacement control approach of constriction at the cap location.
\item Movie M14: Evolution of a axisymmetric constriction profile due to a displacement control approach of constriction at the tube location.
\item Movie M15: Evolution of a non-axisymmetric constriction profile due to a force control approach of constriction at the tube location.
\end{itemize}

\pagebreak

\begin{figure*}[t]
\centering
\includegraphics[width=\linewidth]{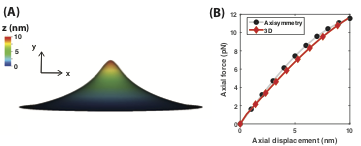}
\caption{Comparison of the axial force required to deform a flat membrane patch up to a height of 10 nm. Bending rigidity is 20 pN$\cdot$nm and membrane tension is 0.1 pN/nm. The results obtained from the axisymmetric model and the 3D framework are compared. The analytical solution for the equilibrium value of force is 12.5664 pN. (A) Membrane shape at a deformation of 10 nm. Colorbar indicates the height (nm). (B) Axial force vs height of membrane in axisymmetry and 3D. See Movie M12 in the supplementary information for the evolution of the membrane deformation.}
\label{fig:pullout_comparison}
\end{figure*}

\begin{figure*}[t]
\centering
\includegraphics[width=0.6\linewidth]{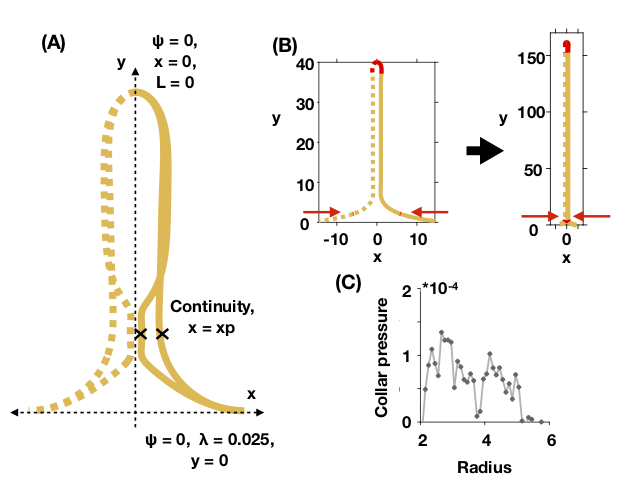}
\caption{No snap-through instability is observed for constriction at the base of a tubule without the fixed height boundary condition. Membrane tension is 0.2 pN/nm, bending rigidity is 320 pN$\cdot$nm. (A) Schematic depicting the boundary conditions used. The difference with the B.Cs in Eq. \ref{BC_tubule} is that the height is no longer constrained. This implies that the axial force is fixed.  Thus, this simulation represents a system of 12 equations with 1 unknown parameter and 13 boundary conditions (Eq. \ref{eq:s-area-systemofequations2}). (B) Initial and final membrane shapes obtained for constriction at the base of the tubule. (C) Collar pressure vs radius at the break point. Pressure is negligible (order of magnitude is 10$^{-4}$.) }
\label{fig:S1_boundary_conditions}
\end{figure*}

\begin{figure*}[t]
\centering
\includegraphics[width=0.8\linewidth]{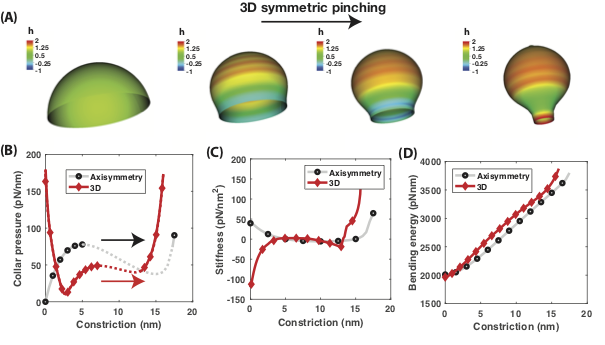}
\caption{Comparison of collar pressure, stiffness and bending energy during constriction of a membrane cap using the axisymmetric and 3D framework. Axisymmetry is enforced in the 3D simulation by solving as a displacement control problem. Boundary conditions used are shown in Fig. 1, case 1. Bending rigidity is 320 pN/nm, Radius is 20 nm.  (A) Membrane shapes during constriction of spherical membrane in 3D. Colorbar is the normalized mean curvature. (B) Collar pressure vs constriction in 3D and axisymmetry. (C) Stiffness vs constriction in 3D and in axisymmetry. (D) Bending energy vs constriction in 3D and axisymmetry. See Movie M13 in the supplementary information for the evolution of the constriction process.}
\label{fig:disp_control_cap_pinching}
\end{figure*}

\begin{figure*}[t]
\centering
\includegraphics[width=0.8\linewidth]{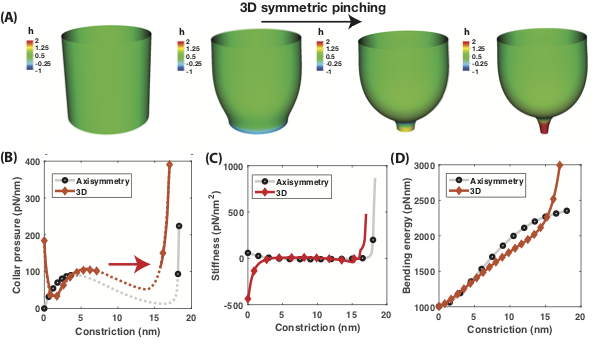}
\caption{Comparison of collar pressure, stiffness and bending energy during constriction of a membrane cylinder using the axisymmetric and 3D framework. Axisymmetry is enforced in the 3D simulation by solving as a displacement control problem. Boundary conditions used are those shown in Fig. 1, case 2. Bending rigidity is 320 pN $\cdot$ nm, length scale R$_0$ is 20 nm. (A) Membrane shapes during constriction of cylindrical membrane in 3D. Colorbar is the normalized mean curvature. (B) Collar pressure vs constriction in 3D and axisymmetry. (C) Stiffness vs constriction in 3D and in axisymmetry. (D) Bending energy vs constriction in 3D and axisymmetry. See Movie M14 in the supplementary information for the evolution of the constriction process.}
\label{fig:disp_control_tube_pinching}
\end{figure*}

\begin{figure*}[t]
\centering
\includegraphics[width=0.8\linewidth]{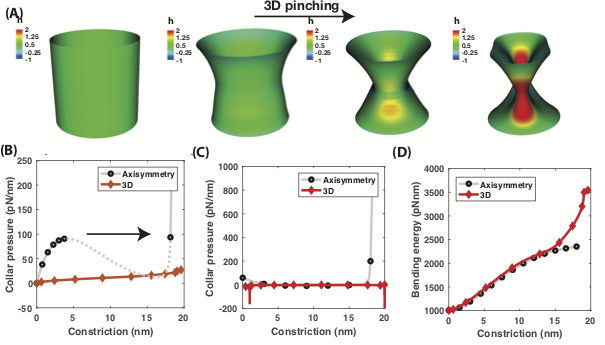}
\caption{Comparison of collar pressure, stiffness and bending energy during constriction of a membrane cylinder using the axisymmetric and 3D framework. Bending rigidity is 320 pN/nm, Radius is 20 nm. (A) Membrane shapes during constriction of cylindrical membrane in 3D. Colorbar is the normalized mean curvature. (B) Collar pressure vs constriction in 3D and axisymmetry. (C) Stiffness vs constriction in 3D and in axisymmetry. (D) Bending energy constriction in 3D and axisymmetry. See Movie M15 in the supplementary information for the evolution of the constriction process.}
\label{fig:force_control_tube_pinching}
\end{figure*}

\begin{figure}[!!h]
\centering
\includegraphics[width=0.8\linewidth]{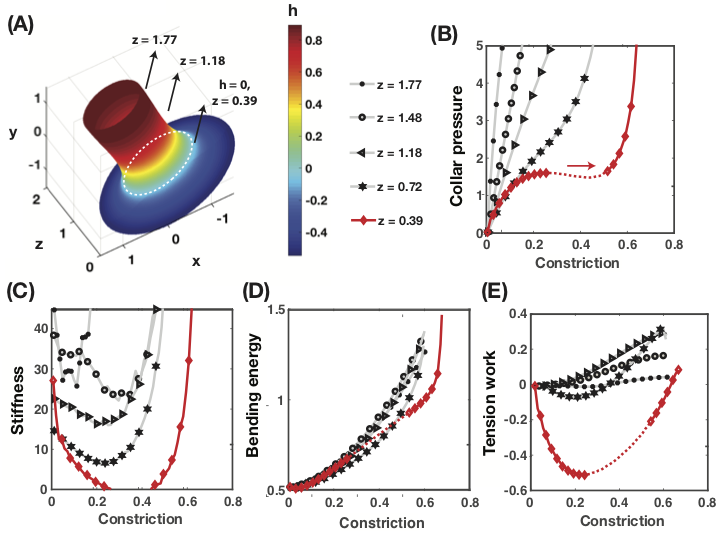}
\caption{The snap-through instability for constriction at the base is regulated by a variation in local tension. Membrane tension at the boundary is 0.2 pN/nm, bending rigidity is 320 pNnm, Radius is 20 nm, area of applied force is 1/200th of the membrane area. z is the non-dimensional height at a given location along the membrane from the bottom. Shown are the (A) Mean curvature distribution (non-dimensional) and the location of the local minimal surface (dotted line at y = 0.39) where the mean curvature vanishes ($h = 0$), (B) Collar pressure, (C) Tubule stiffness to pinching, (D) Bending energy and (E) Tension work as a function of the constriction.  }
\label{fig:fig3_axisymmetric_base}
\end{figure}

\end{document}

%% file: mathSymbols.tex
\def\ba{\mbox{\boldmath$ a$}}

\def\be{\mbox{\boldmath$ e$}}

\def\bk{\mbox{\boldmath$ k$}}

\def\bn{\mbox{\boldmath$ n$}}

\def\br{\mbox{\boldmath$ r$}}
\def\bs{\mbox{\boldmath$ s$}}

\def\bu{\mbox{\boldmath$ u$}}